\documentstyle[emulateapj,psfig,natbib,amsfonts,amsmath]{article}
\def\ra#1#2#3{#1$^{\rm h}$#2$^{\rm m}$#3$^{\rm s}$}
\def\dec#1#2#3{$#1^\circ#2'#3''$}
\def\swift{{\it Swift}}
\def\nod{\nodata}
\def\grb{GRB\,050505}

\def\ociw{1}
\def\prince{2}
\def\hubble{3}
\def\pom{4}
\def\srl{5}
\def\cit{6}
\def\psu{7}

\begin{document}

\title{Spectroscopy of GRB\,050505 at $z=4.275$: A ${\rm log}\,N({\rm 
HI})=22.1$ DLA Host Galaxy and the Nature of the Progenitor}

\author{
E.~Berger\altaffilmark{\ociw,}\altaffilmark{\prince,}\altaffilmark{\hubble},
B.~E.~Penprase\altaffilmark{\pom},
S.~B.~Cenko\altaffilmark{\srl},
S.~R.~Kulkarni\altaffilmark{\cit},
D.~B.~Fox\altaffilmark{\psu},
C.~C.~Steidel\altaffilmark{\cit},
and N.~A.~Reddy\altaffilmark{\cit}
}

\altaffiltext{\ociw}{Observatories of the Carnegie Institution
of Washington, 813 Santa Barbara Street, Pasadena, CA 91101}
 
\altaffiltext{\prince}{Princeton University Observatory,
Peyton Hall, Ivy Lane, Princeton, NJ 08544}
 
\altaffiltext{\hubble}{Hubble Fellow}

\altaffiltext{\pom}{Pomona College Department of Physics and Astronomy,
610 N. College Avenue, Claremont, CA}

\altaffiltext{\srl}{Space Radiation Laboratory, MS 220-47, California
Institute of Technology, Pasadena, CA 91125}

\altaffiltext{\cit}{Division of Physics, Mathematics and Astronomy,
105-24, California Institute of Technology, Pasadena, CA 91125}

\altaffiltext{\psu}{Department of Astronomy and Astrophysics,
Pennsylvania State University, 525 Davey Laboratory, University
Park, PA 16802}

\begin{abstract}
We present the discovery of the optical afterglow of \grb\ and an
optical absorption spectrum obtained with the Keck I 10-m telescope.
The spectrum exhibits three redshifted absorption systems with the
highest, at $z=4.2748$, arising in the GRB host galaxy.  The host
absorption system is marked by a damped Ly$\alpha$ (DLA) feature with
a neutral hydrogen column density of ${\rm log}\,N({\rm HI})=22.05\pm
0.10$, higher than that of any QSO-DLA detected to date, but similar
to several other recent measurements from GRB spectra.  In addition,
we detect absorption lines from both low- and high-ionization species
from which we deduce a metallicity, $Z\approx 0.06$ $Z_\odot$, with a
depletion pattern that is roughly similar to that of the Galactic warm
halo, warm disk, or disk+halo.  More importantly, we detect strong
absorption from \ion{Si}{2}* indicating a dense environment, $n_H
\gtrsim 10^2$ cm$^{-3}$, in the vicinity of the burst, with a size of
$\sim 4$ pc.  In addition, the \ion{C}{4} absorption system spans a
velocity range of about $10^3$ km s$^{-1}$, which is not detected in
any other absorption feature.  We show that the most likely
interpretation for this wide velocity range is absorption in the wind
from the progenitor star.  In this context, the lack of corresponding
\ion{Si}{4} absorption indicates that the progenitor had a mass of 
$\lesssim 25$ M$_\odot$ and a metallicity $\lesssim 0.1$ $Z_\odot$,
and therefore required a binary companion to eject its hydrogen
envelope prior to the GRB explosion.  Finally, by extending the
GRB-DLA sample to $z\approx 4.3$ we show that these objects appear to
follow a similar metallicity-redshift relation as in QSO-DLAs, but
with systematically higher metallicities.  It remains to be seen
whether this trend is simply due to the higher neutral hydrogen
columns in GRB-DLAs, or if it is a manifestation of different star
formation properties in GRB-DLAs.  Thus, GRBs hold the potential to
probe all scales relevant to our understanding of the star formation
process and its relation to metal production.
\end{abstract}
 
\keywords{gamma-rays:bursts --- ISM: abundances --- ISM:kinematics 
---  stars: mass loss --- stars: Wolf-Rayet}

\section{Introduction}
\label{sec:intro}

The study of star formation and the associated production and
enrichment of the interstellar medium (ISM) and intergalactic medium
(IGM) by metals is focused on two observational approaches: large
galaxy surveys based on rest-frame UV, optical, near-infrared and
far-infrared emission, and the use of high redshift quasars as probes
of the ISM/IGM.  The former provides a view of the star formation
activity from high redshift to the present, as well as the possible
enrichment of the IGM through galactic-scale winds.  In the context of
quasar absorption studies the highest column density absorbers, the
so-called damped Lyman alpha (DLA) systems ($N({\rm HI})\geq 2\times
10^{20}$ cm$^{-2}$) are of particular interest.  This is because of
the similarity of their column densities to those in local luminous
galaxies, the fact that at these columns the hydrogen is mainly
neutral and may form stars, and the conclusion that the bulk of the
neutral gas in the range $z\sim 0-5$ is in DLAs (for a recent review
see \citealt{wgp05} and references therein).

The connection between luminous star-forming galaxies and DLAs remains
an open question.  This is primarily due to observational limitations
inherent in both approaches.  First, surveys of star-forming galaxies
at $z\gtrsim 2$ are limited to relatively bright object, $L\gtrsim
0.3L*$ (e.g., \citealt{sas+03}).  Furthermore, with the exception of
the brightest galaxies, the continuum emission is typically too faint
to observe absorption from interstellar gas and therefore to tie
together the star formation and ISM properties.  On the other hand,
the use of quasar sight lines to probe the disks of high redshift
galaxies is limited by a cross-section selection effect to large
impact parameters ($\gtrsim 10$ kpc).  Moreover, in the case of DLAs
the small impact parameters severely limit the identification of the
DLA counterparts against the bright background quasar.  Thus, while
some DLA counterparts have been identified, primarily at $z\lesssim 1$
(e.g., \citealt{cl03}), a definitive association with star-forming
systems or LBGs remains elusive.

Despite the observational challenges, the various galaxy surveys have
provided evidence for near solar metallicities in at least some of the
bright high redshift galaxies \citep{sep+04,ssc+04}.  Typical star
formation rates in the bright galaxies are in the range of $\sim
10-10^2$ M$_\odot$ yr$^{-1}$ \citep{ssa+01,fdf+04,ssc+04}.  On the
other hand, DLAs appear to be metal poor, with a typical $Z\sim 0.03$
$Z_\odot$ \citep{pgw+03}, and while they exhibit evidence for star
formation in a few cases, the rates typically appear to be lower than
in the LBGs \citep{bwc+99}.

An alternative approach to probing intervening gas in galaxies and the
IGM is to use the afterglows of gamma-ray bursts (GRBs).  In the
context of the relation to star formation and the nature of DLAs, GRBs
offer several advantages over quasar studies.  First, GRBs are
embedded in star forming galaxies with typical offsets of a few kpc or
less \citep{bkd02}.  They therefore not only provide a direct link to
star formation, but also probe the regions of most intense star
formation, and hence production and dispersal of metals.  Second,
since the GRB afterglow emission fades away on a timescale of days to
weeks, the host galaxy and any intervening DLAs can be subsequently
studied directly (e.g., \citealt{vel+04}).

Third, and perhaps most important, since GRBs are likely to be located
in star forming regions (e.g., molecular clouds) within their host
galaxies, this approach provides the only systematic way to directly
probe the small-scale environment and conditions of star formation at
high redshift; the probability of intersecting an individual molecular
cloud in a quasar sight line is vanishingly small.  In the same vein,
GRBs can probe the circumstellar environment of the progenitor star
itself and provide a unique view of the mass loss history and
properties of the progenitor (e.g., metallicity, mass, binarity).
With a large statistical sample, this is the only way to compare the
properties of massive stars and individual star forming regions at
high redshift to those in the Milky Way and the local universe.

Over the past several years, a few absorption spectra of GRB
afterglows have been obtained, revealing relatively large neutral
hydrogen column densities (most in the DLA category; e.g.,
\citealt{vel+04}).  The metallicity, inferred in only a few 
cases, appears to be sub-solar ($Z\sim 0.01-0.1$ $Z_\odot$;
\citealt{vel+04,cpb+05,sve+05}), but with a dust-to-gas ratio that 
is larger than that in QSO-DLAs \citep{sff03}.  In addition, some
spectra reveal complex velocity structure, interpreted to arise from
ordered galactic rotation \citep{cgh+03}, and in one case appearing to
arise in the complex wind environment of the progenitor star (e.g.,
\citealt{mfh+02}).  These results already suggest that GRBs 
probe environments that are likely missed in the quasar surveys.

With the advent of the \swift\ satellite, we can now start to utilize
GRBs as probes of the high redshift universe in a systematic manner.
\swift's sensitivity and ability to rapidly and accurately localize a
large number of GRBs has resulted in a redshift distribution spanning
nearly uniformly from $z\sim 0.5$ to $\gtrsim 6$
\citep{bkf+05,jlf+05,kyk+05,hnr+05}.  This sample is therefore 
well-matched to the star formation history of the universe, and over
time will allow us to address the redshift evolution of star forming
environments and perhaps individual massive stars.

Along these lines, we present here an absorption spectrum of \grb,
which reveals a DLA with a column density ${\rm log}\,N({\rm HI})=22.05
\pm 0.10$ at a redshift of $z=4.2748$.  This system is currently the 
highest redshift GRB host for which detailed information is available.
The spectrum probes not only the interstellar medium of the host
galaxy, but also provides information on the local environment of the
burst, likely including the wind of the progenitor star from which we
are able to draw conclusions about the nature of the star that
exploded.

\section{Observations}
\label{sec:obs}

\grb\ was detected by \swift\ on 2005 May 5.974 UT.  The duration and 
fluence of the burst are 60 s and $(4.1\pm 0.4)\times 10^{-6}$ erg
cm$^{-2}$, respectively \citep{gcn3364}.  Observations with the
\swift\ X-ray telescope (XRT) started 47 min after the burst, and
revealed an uncatalogued source at $\alpha$=\ra{09}{27}{03.2},
$\delta$=\dec{+30}{16}{21.5} (J2000) with an uncertainty of about
$6''$ and a flux of $\sim 2\times 10^{-11}$ erg cm$^{-2}$ s$^{-1}$
\citep{gcn3365} .  No object was detected by the \swift\ UV/optical
telescope (UVOT) in the first 8 hours to a limit of $V>20.35$ mag and
$B>21.04$ mag, at a mean time of 2.49 and 2.59 hr, respectively
\citep{gcn3371}.

\begin{figure}
\centerline{\psfig{file=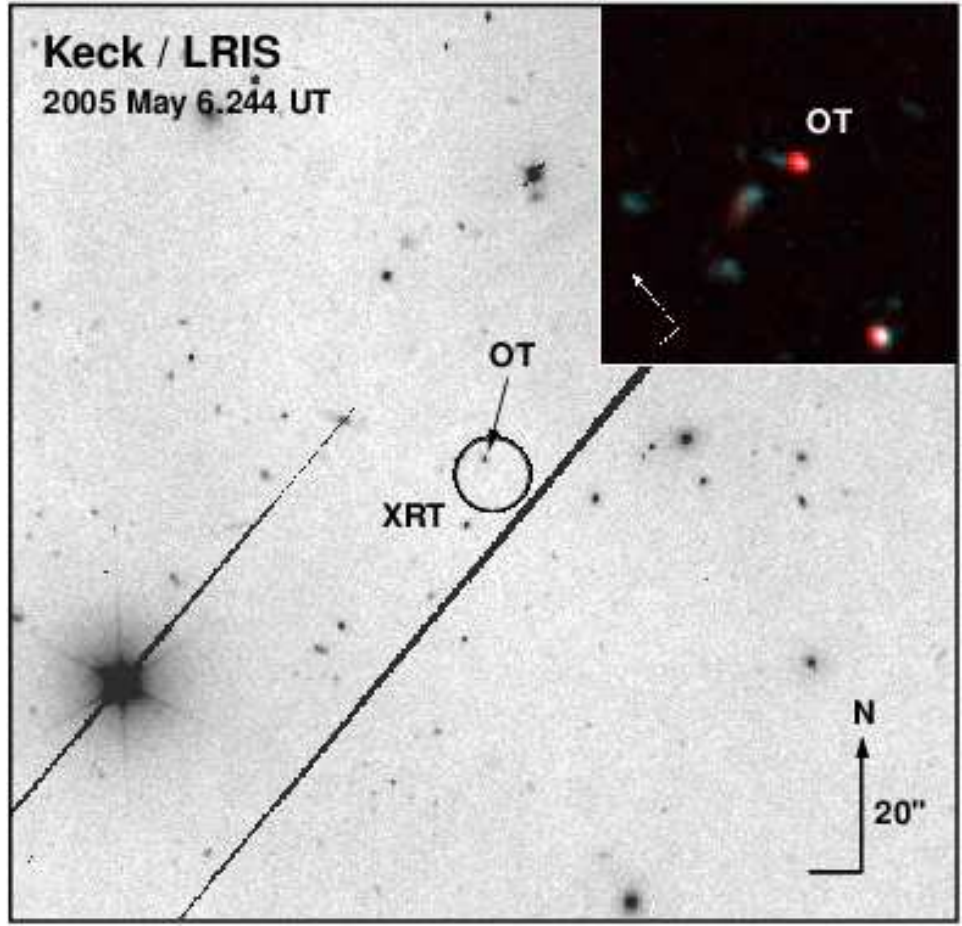,width=5.0in}}
\caption{Discovery image of the optical afterglow of \grb\ obtained
with LRIS on the Keck I 10-m telescope.  The inset is a combined $g+I$
flux-calibrated color image clearly showing the red color of the
afterglow which is due to the damped Ly$\alpha$ absorption, the
Ly$\alpha$ forest, and the Lyman limit absorption at $\lambda\lesssim
4900$\AA\ (see Figure~\ref{fig:flam}).
\label{fig:lris}}
\end{figure}

We initiated observations of \grb\ with the Low Resolution Imaging
Spectrometer (LRIS) mounted on the Keck I 10-m telescope about 6.4 hr
after the burst.  We obtained simultaneous $g$- and $I$-band
observations and detected an object close to the center of the XRT
error circle at $\alpha$=\ra{09}{27}{03.3},
$\delta$=\dec{+30}{16}{23.7} (J2000), with a brightness $I=20.51\pm
0.05$ mag and $g=23.67\pm 0.12$ mag (Figure~\ref{fig:lris}).  The
spectral slope between the two bands, $F_\nu\propto \nu^{-4.9\pm
0.3}$, is too sharp for host galaxy extinction, and instead suggests a
redshift $z\sim 4-5.5$.  Contemporaneous observations with UKIRT
revealed a near-IR counterpart with a brightness of $K=18.1\pm 0.2$
mag \citep{gcn3372}.

Following the identification of the afterglow we used LRIS to obtain
two 900-s spectra with a $1''$ wide slit.  The spectra were reduced
using standard IRAF routines, while rectification and sky subtraction
were performed using the method of \citet{kel03}.  Wavelength
calibration was performed using HgArNeZnCd arc lamps and air-to-vacuum
and heliocentric corrections were applied.  The resulting dispersion
scales are 1.86 \AA/pix for the red side and 0.61 \AA/pix for the blue
side, with an rms wavelength uncertainty of about 0.2\AA.  Finally,
flux calibration was performed using the spectrophotometric standard
BD+284211, with a correction for the small amount of Galactic
extinction, $E(B-V)=0.021$ mag \citep{sfd98}.  Convolution of the
spectrum with the $I$ filter bandpass results in a flux density of
$15.8\pm 1.5$ $\mu$Jy ($20.5\pm 0.1$ mag), in very good agreement with
the measured $I$-band magnitude.  This indicates that slit losses were
minimal.

We identify three redshift systems in the extracted spectrum, $z_1=
4.2748\pm 0.0008$, $z_2=2.2650\pm 0.0008$, and $z_3=1.6948\pm 0.0003$.
We associate the redshift system $z_1$ with the host galaxy of \grb\
given the nature of the absorption lines (see \S\ref{sec:abs}) and the
lack of Ly$\alpha$ forest absorption redward of the damped Ly$\alpha$
feature at the redshift $z_1$.  At this redshift, using the standard
cosmology ($H_0=71$ km s$^{-1}$ Mpc$^{-1}$, $\Omega_m=0.27$ and
$\Omega_\Lambda=0.73$), the isotropic-equivalent $\gamma$-ray energy
release is $E_{\rm \gamma,iso}\approx 8.9\times 10^{53}$ erg, while
the X-ray luminosity extrapolated to $t=10$ hr (assuming the typical
$F_\nu\propto t^{-1.3}$) is $L_{X,{\rm iso}}\approx 2.3\times 10^{46}$
erg s$^{-1}$ \citep{nkg+05}.  Both values are at the high end of the
distribution for previous GRBs \citep{bkf03,bfk03}.

\section{The Host Galaxy Absorption System}
\label{sec:abs}

The spectrum of \grb\ is dominated by a broad absorption feature
centered on an observed wavelength of about 6400\AA, which we identify
as Ly$\alpha$.  A Voigt profile fit to the Ly$\alpha$ absorption
feature results in a column density of neutral hydrogen, ${\rm
log}\,N({\rm HI})= 22.05\pm 0.10$, indicating that the absorber is a
DLA, with one of the highest column densities measured to date in
either QSO or GRB sight lines \citep{cwm+02,vel+04}.  We note that the
actual column density may be somewhat higher since, unlike in the case
of quasars, the GRB is embedded in the host galaxy
\citep{vel+04}.  The average flux decrement between Ly$\alpha$ and
Ly$\beta$ is about 0.55, which is in good agreement with the values
measured in quasar sight lines at the same redshift \citep{cgb+93}.

Blueward of the damped Ly$\alpha$ feature we detect absorption from
the Ly$\alpha$ forest, likely mixed with metal absorption features of
oxygen, silicon, nitrogen, sulfur, carbon, and iron.  Due to the low
resolution of our spectrum we cannot disentangle the contribution of
the metals from the forest.  We also detect Ly$\beta$ absorption, and
at a wavelength of about 4880\AA\ we detect the Lyman limit.

Redward of the Ly$\alpha$ absorption feature we detect a wide range of
metal lines.  The identifications, observed wavelengths, and
equivalent widths (EWs) of these lines are listed in
Table~\ref{tab:lines}.  We note that in some cases these lines are
blended with absorption features from the lower redshift system at
$z_2=2.265$.  Most of the lines detected in the spectrum appear to be
saturated.  We can obtain a lower limit on the column density assuming
the optically thin case
\begin{equation}
N=\frac{m_ec^2}{\pi e^2}\frac{W_\lambda}{f\lambda^2} 
=1.13\times 10^{20}\,{\rm cm}^{-2}\, \frac{(W_\lambda/{\rm \AA})}
{(\lambda/{\rm \AA})^2f},
\label{eqn:thin}
\end{equation}
where $f$ is the oscillator strength, $W_\lambda$ is the equivalent
width, and $\lambda$ is the rest wavelength.  The resulting lower
limits for each transition are listed in Table~\ref{tab:lines}.

\begin{figure}
\centerline{\psfig{file=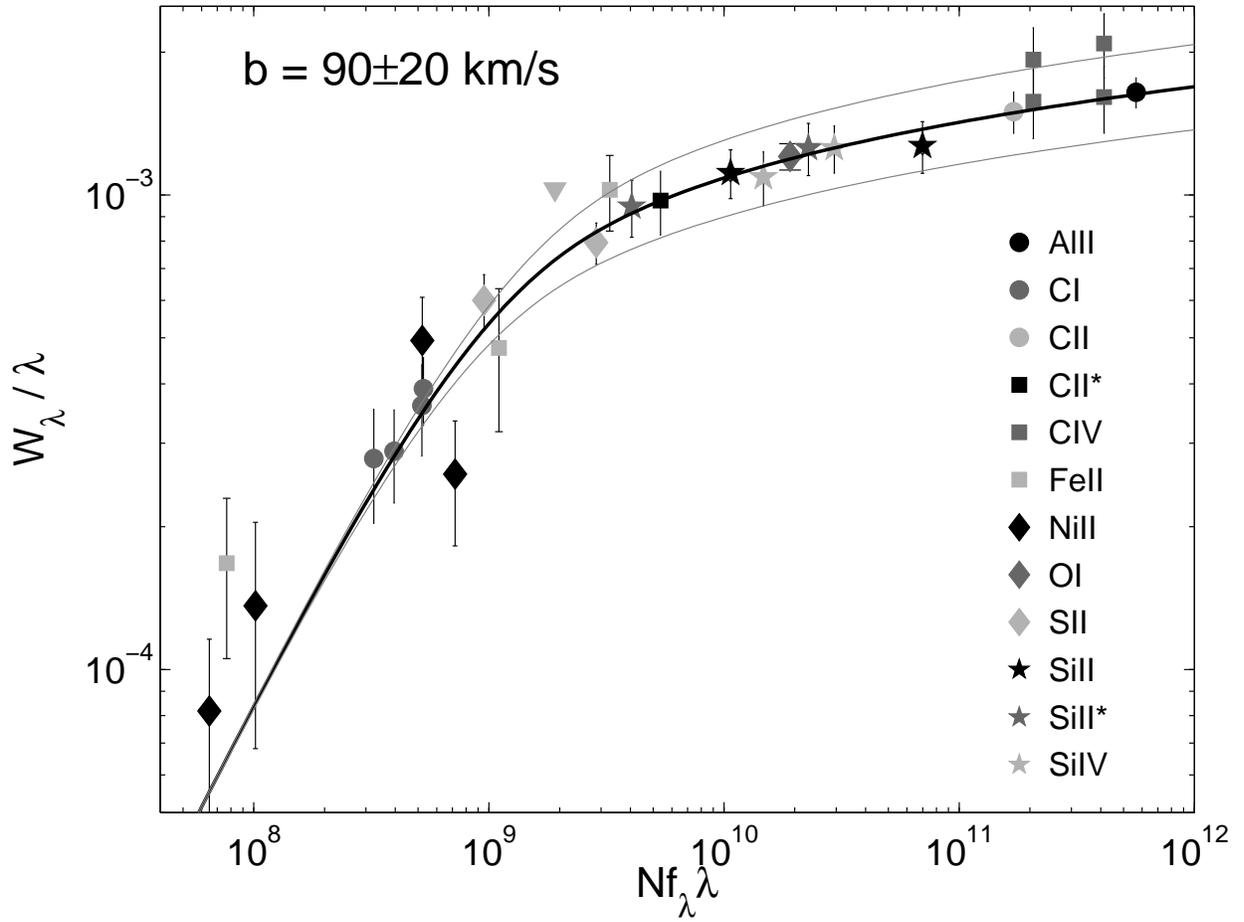,width=6.5in}}
\caption{Curve of growth (COG) for the host galaxy system of \grb.  
We constructed the COG by iteratively fitting for the column densities
of individual ions and the Doppler parameter, $b$, which we assumed to
have a single value.  Transitions in the linear part of the COG lead
to well-determined columns (modulo the low resolution of our
spectrum).  Transitions on the flat portion of the COG are sensitive
to the value of $b$ and should be considered as lower limits.
\label{fig:cog}}
\end{figure}

Since most of the lines are saturated and therefore do not lie on the
linear part of the curve of growth (COG) we construct a joint COG for
all the detected transitions which do not suffer from strong blending.
In this case we use the standard formulation \citep{spi78}
\begin{equation}
W_\lambda = \frac{2bF(\tau_0)\lambda}{c}
\end{equation}
where the line center's optical depth is given by
\begin{equation}
\tau_0=\frac{\pi^{1/2}e^2f\lambda N}{m_ecb}
=1.496\times 10^{-15}\frac{f(\lambda/{\rm \AA})N}{(b/{\rm km\,s^{-1}})}
\end{equation}
and the function 
\begin{equation}
F(\tau_0) = \int_{0}^{\infty}[1-{\rm exp}(-\tau_0{\rm e}^{-x^2})]{\rm d}x.
\end{equation}
We assume that all species share the same value of the Doppler
parameter, $b$, and we fit iteratively for $b$ and the column density
of each species (e.g., \citealt{sff03}).  The resulting best-fit COG
is shown in Figure~\ref{fig:cog}, and the column densities are listed
in Table~\ref{tab:columns}.  We find that lines of \ion{C}{1} and
\ion{Ni}{2} are well-described by the linear part of the COG, lines of
\ion{S}{2} are mildly saturated, and most other species lie on the 
flat portion of the COG.  We stress that given the assumption of a
single $b$ value, as well as the low resolution of our spectrum, the
derived column densities of the strongly saturated lines should still
be considered as rough lower limits.

With this caveat in mind we derive the following abundances from the
COG analysis.  The column density of \ion{S}{2} is ${\rm
log}\,N\approx 16.1$, leading to an abundance relative to the solar
value of ${\rm [S/H]}\approx -1.2$; here we use the compilation of
solar abundances of \citet{ags05}.  Sulfur is a non-refractory element
and its gas-phase abundance therefore closely matches the gas
metallicity.  The derived value is roughly similar to those of both
QSO- and GRB-DLAs.  The \ion{Fe}{2} column density of ${\rm
log}\,N\approx 15.5$ indicates an abundance of ${\rm [Fe/H]}\approx
-2.0$, which is similar to the mean value for QSO-DLAs at a similar
redshift range (e.g., \citealt{pgw+03}).  The ratio of sulfur to iron,
${\rm [S/Fe]}\approx 0.8$, however, is at the high end of values
measured for QSO-DLAs, which at ${\rm [Fe/H]}\approx -2.0$ range from
about 0.1 to 1 \citep{le03}.  Since iron is a strongly depleted
element, the large value of [S/Fe] suggests a high dust content,
similar to the inference made in several other GRB-DLAs
\citep{sff03,sf04,wfl+05}.

Inspecting the abundances of other ions we find that ${\rm [Si/H]}
\approx -1.6$, when we sum the contributions of \ion{Si}{2} and
\ion{Si}{4}.  The ratio compared to iron is ${\rm [Si/Fe]}\approx 
0.4$, which is in excellent agreement with the median value of about
0.4 for QSO-DLAs \citep{wgp05}.  

\begin{figure}
\centerline{\psfig{file=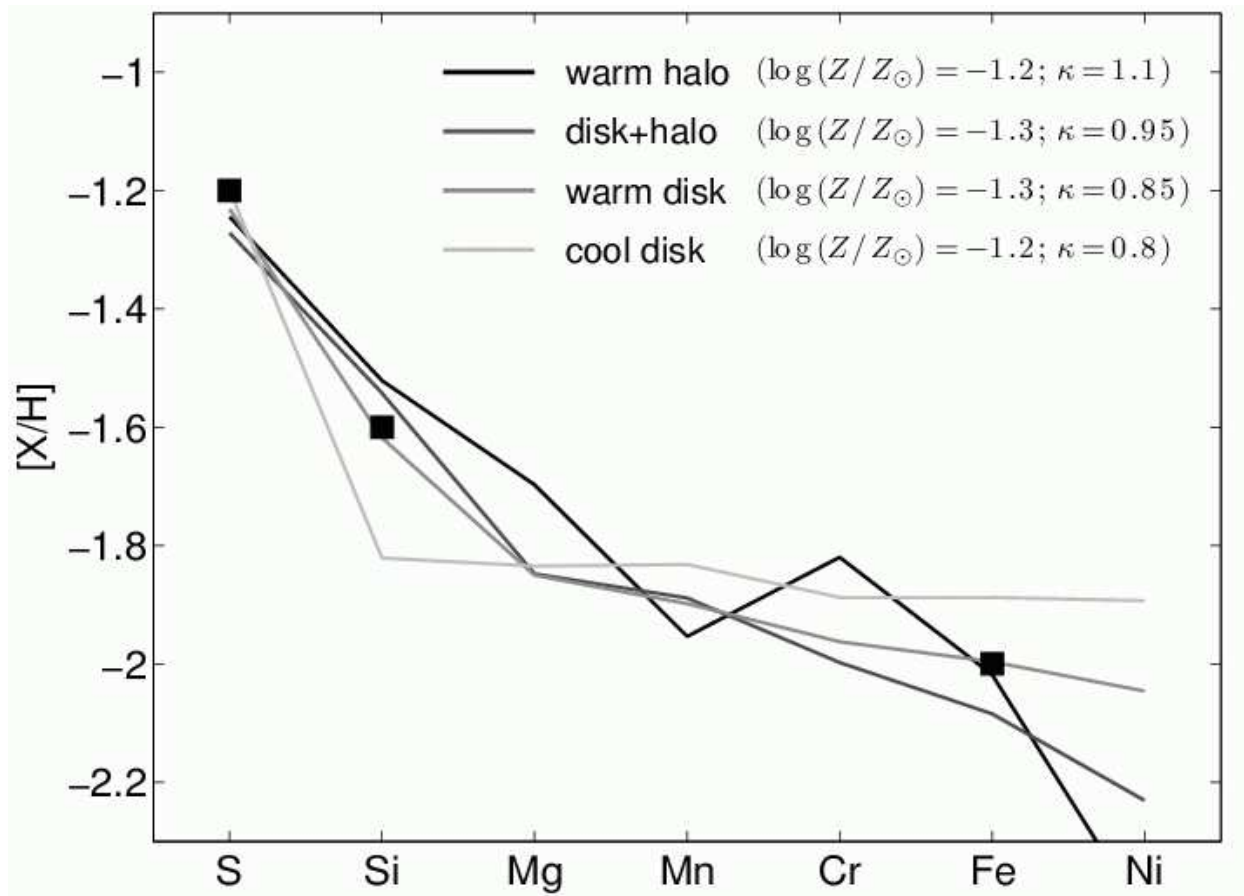,width=6.5in}}
\caption{Depletion pattern for the host galaxy of \grb\ using the 
inferred columns of sulfur, silicon, and iron relative to hydrogen.
The lines represent the depletion patterns expected for various phases
of the interstellar medium based on observations in the Milky Way
\citep{ss96}, with the dust-to-gas ($\kappa$) and metallicity left as 
free parameters.  We find that a cool disk interpretation is unlikely.
\label{fig:dep}}
\end{figure}

Using the abundances of sulfur, silicon, and iron we provide a
comparison to the Milky Way depletion patterns in
Figure~\ref{fig:dep}.  We consider the four typical patterns of warm
halo (WH), warm disk+halo (WDH), warm disk (WD), and cool disk (CD)
clouds \citep{ss96}, and follow the method of \citet{sff03} to
determine the metallicity and dust-to-gas ratio in each case.  We find
that the first three depletion patterns provide an adequate
representation of the data with a metallicity $Z\approx 0.06$
$Z_\odot$ and dust-to-gas ratios of 1.1 (WH), 0.95 (WDH), and 0.85
(WD).  The cool disk depletion pattern does not provide an adequate
fit since it underestimates the silicon abundance and over-estimates
the iron abundance.  Given the inferred metallicities and dust-to-gas
ratios we calculate a rest-frame $V$-band extinction of about $0.3\pm
0.1$ mag.

\subsection{A \ion{C}{4} Outflow}

\begin{figure}
\centerline{\psfig{file=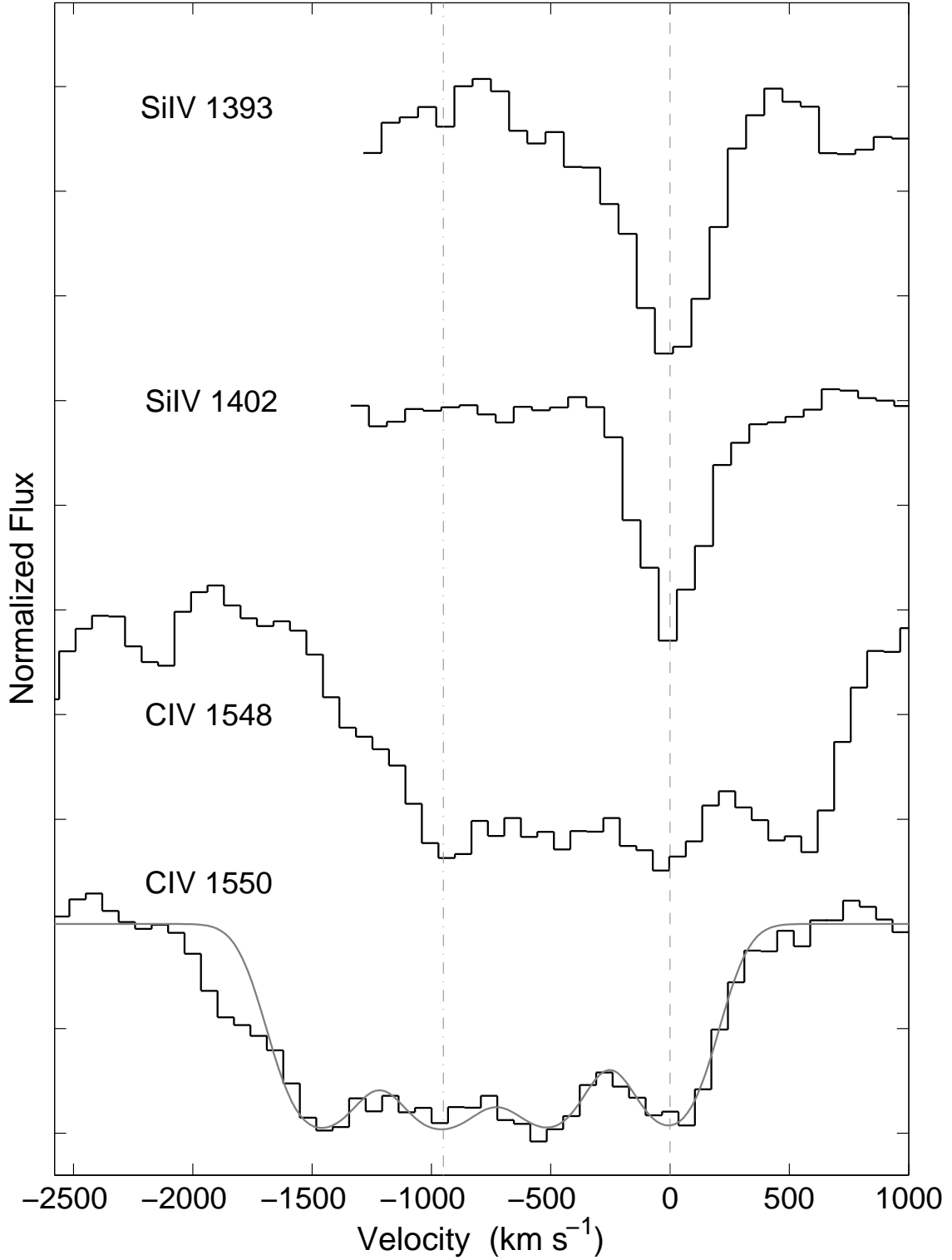,width=6.0in}}
\caption{Absorption features of \ion{C}{4} and \ion{Si}{4}.  The 
\ion{C}{4} doublet exhibits a velocity spread of about 1000 km 
s$^{-1}$, which is not detected in \ion{Si}{4}.  The gray line shows
the expected absorption profile from two systems at $z_{1A}=4.2741$
and $z_{1B}=4.2572$ with \ion{C}{4} column densities of $1\times
10^{15}$ and $2\times 10^{15}$ cm$^{-2}$, respectively, and $b=90$ km
s$^{-1}$ (see Figure~\ref{fig:cog}).  This scenario provides an
adequate fit to the data, and indicates that the velocity structure is
most likely due to a fast wind from the progenitor star.
\label{fig:civ-siiv}}
\end{figure}

The most prominent metal absorption feature in the spectrum of \grb,
with a total rest-frame equivalent width of about 11.6\AA, is a blend
of \ion{C}{4}$\lambda\lambda\,1548,1550$.  The overall velocity spread
of this feature is about 950 km s$^{-1}$, extending blueward of the
systemic redshift of the host galaxy, as defined by other metal
absorption features (Figure~\ref{fig:civ-siiv}).  The wide velocity
spread is not observed in any of the other low- or high-ionization
features; for example the limit on blue-shifted \ion{Si}{4} is ${\rm
log}\,N<13.5$, at least a factor of thirty below the \ion{C}{4} column
density.  It is not clear from the low resolution spectrum if the
velocity structure is due to a set of discrete absorbers or to a
relatively uniform distribution of \ion{C}{4}.

Regardless of this distinction, the observed velocity range is much
larger than the typical values associated with the rotation or
velocity dispersion of individual galaxies, including those observed
in past GRB absorption spectra (e.g., GRB\,000926 with $v\approx 170$
km s$^{-1}$; \citealt{cgh+03}).  Similarly, blueshifted absorption by
interstellar low- and high-ionization gas has been detected in the
stacked spectra of Lyman break galaxies (LBGs), but with a typical
velocity shift relative to the nebular emission lines of only $\sim
-150$ km s$^{-1}$ \citep{ass+03}.  On the other hand, \citet{fbb02}
find evidence for outflows in $z\gtrsim 4$ lensed galaxies which
approach 800 km s$^{-1}$, but these outflows are evident in
low-ionization lines of \ion{O}{1}, \ion{Si}{2}, and \ion{C}{2}.  In
local bright starburst galaxies outflow velocities are typically
$\lesssim 300$ km s$^{-1}$ \citep{hls+00}.  Therefore, unless the wind
from the host galaxy of \grb\ is unusually fast, we consider this
interpretation unlikely.  This would particularly be the case if the
host of \grb\ is similar to those of other GRBs, which tend to have
relatively low masses \citep{chg04} and would therefore have escape
velocities much smaller than $10^3$ km s$^{-1}$.

The \ion{C}{4} absorption may be alternatively related to an extended
large-scale structure along the line of sight.  \citet{ass+03} find a
strong correlation between the location of \ion{C}{4} absorbers in
quasar sight lines and the location of LBGs.  In particular, for a
\ion{C}{4} column density of $>10^{14}$ cm$^{-2}$ (as observed in the
spectrum of \grb) about $80\%$ of the absorbers lie within $600$ km
s$^{-1}$ and $\Delta\theta =35''$ of an LBG.  While this velocity
range is similar to the one observed here, it has been argued that the
actual outflow velocities are $\lesssim 400$ km s$^{-1}$
\citep{son05}, and that the extent of the \ion{C}{4} bubbles may be a
reflection of pre-galactic enrichment at $z\sim 10$ by slower winds
from dwarf galaxies \citep{pm05}.  Since here we directly detect an
outflow with $v\sim 10^3$ km s$^{-1}$ it is unlikely to be related to
the \ion{C}{4} bubbles discussed in \citet{ass+03} and \citet{son05}.

Another possibility in the context of an extended structure is that
the \ion{C}{4} absorption arises from an overlapping structure of
galactic halos, perhaps stretched out along a dark matter filament.
This hypothesis is difficult to assess with a single absorption
spectrum.  However, if such structures are ubiquitous they should be
observed in a large fraction of GRB absorption spectra, as well as in
quasar absorption spectra.  Studies of \ion{C}{4} systems in quasar
spectra suggest that these systems are strongly clustered for velocity
shifts of $\lesssim 200$ km s$^{-1}$ and are essentially uncorrelated
for $v\gtrsim 500$ km s$^{-1}$ \citep{rsw+96,psa+03}.  Thus, we
conclude that the observed \ion{C}{4} velocity structure is not due to
a galactic-scale phenomenon.

In the context of a massive star progenitor, an attractive alternative
is that the \ion{C}{4} velocity structure is the signature of a fast
wind.  A similar inference was made in the case of GRB\,021004, which
exhibited a complex velocity structure extending from about 140 to
3000 km s$^{-1}$ \citep{mfh+02,mhc+03,sgh+03,fdl+05,swh+05}.  The
presence of such a fast outflow led to the general conclusion that the
progenitor was a Wolf-Rayet star.  However, unlike in the case
presented here, the velocity structure in the spectrum of GRB\,021004
was evident in both low- and high-ionization lines, including
\ion{H}{1}.  This raised the difficulty of explaining the presence of
both types of species in the highly-ionized burst environment, as well
as the presence of a large column of hydrogen in a Wolf-Rayet wind.

While these difficulties prevented a definitive explanation of the
high-velocity absorbers, two main possibilities have been
discussed. \citet{mhc+03} argue that to reach the observed expansion
velocities, radiative acceleration of the pre-existing fragmented
shell nebula is required.  In this model the interaction of the fast
Wolf-Rayet wind with the slower winds of earlier mass loss episodes
results in a fragmented structure due to Rayleigh-Taylor
instabilities.  The radiation pressure produced by the GRB and
afterglow emission then accelerates the fragmented shell nebula to
velocities of hundreds to thousands of km s$^{-1}$.

\citet{mlg05}, on the other hand, produce hydrodynamic numerical 
simulations of the interactions of various mass-loss phases (fast main
sequence wind, slow red supergiant wind, and a fast wolf-Rayet wind),
and conclude that these interactions lead to a clumpy structure with a
velocity range in excess of 2000 km s$^{-1}$.  The fragmented and
clumpy structure of the wind is a common theme in both scenarios, and
is used to explain the similar kinematic structure in both the low-
and high-ionization lines.

In the case of \grb\ the velocity structure is only apparent in
\ion{C}{4}, suggesting that a more uniform wind structure is allowed.  
Still, the effect of the ionizing radiation of the GRB has to be taken
into account.  Since the recombination timescale is expected to be
significantly longer than the lifetime of the GRB (unless the density
is $\gtrsim 10^6$ cm$^{-3}$; \citealt{pl98}), we can estimate the size
of the ionized region roughly as $r\sim \sqrt{E_{\rm ion}\sigma_{\rm
CIV}/4\pi e_{\rm ion,CIV}}\sim {\rm few}\times 10^{19}$ cm; here
$E_{\rm ion}$ is the total energy in photons that can ionize
\ion{C}{4}, $\sigma_{\rm CIV}=1.068\times 10^{-16}$ cm$^{2}$ is the
photoionization cross-section, and $e_{\rm ion,CIV}=64.5$ eV is the
ionization threshold energy.  Clearly, a detailed photoionization
calculation is required, but we conclude that outside of a few parsec
\ion{C}{4} ions should remain intact.  We stress that the minimum
radius for survival of \ion{Si}{4} ions is a factor of about 3.5 times
smaller than that of \ion{C}{4}, so that the non-detection of
outflowing \ion{Si}{4} cannot be simply attributed to photoionization.

Thus, the observations require a fast-moving ($\sim 10^3$ km s$^{-1}$)
wind, enriched in carbon and deficient in silicon, and with a radius
of at least a few parsec.  These requirements can be naturally
satisfied in the case of a Wolf-Rayet wind for which typical
velocities can easily exceed $10^3$ km s$^{-1}$ and extend out to a
distance of $\sim 20$ pc \citep{abb78,mlg05}.

The lack of silicon absorption, with a column density that is at least
a factor of thirty lower than that of \ion{C}{4}, provides interesting
constraints on the nature of the progenitor.  Calculations of
theoretical profiles of \ion{C}{4} and \ion{Si}{4} in the winds from a
range of massive stars suggest that while strong \ion{C}{4} absorption
is nearly always present, the presence of \ion{Si}{4} lines depends
sensitively on a star's mass and metallicity \citep{ll91}.  Thus, the
wind from progenitors with a low mass and metallicity will have
negligible \ion{Si}{4} absorption.  For example, in the case of a wind
from a 25 M$_\odot$ star at the end of core hydrogen burning the ratio
of \ion{Si}{4} to \ion{C}{4} matches the observed limit of $EW({\rm
SiIV}) /EW({\rm CIV})<0.1$ for $Z\lesssim 0.1$ $Z_\odot$, while for a
60 M$_\odot$ star the theoretical ratio is $\sim 0.5$ even for
$Z=0.03$ $Z_\odot$.  These considerations indicate that the progenitor
of \grb\ was most likely a carbon-rich (WC) Wolf-Rayet star with a
relatively low mass ($\lesssim 25$ M$_\odot$) and metallicity
($Z\lesssim 0.1$ $Z_\odot$).

However, radiatively-driven winds follow a mass loss rate metallicity
relation, $\dot{M}\propto Z^{0.5-0.7}$ \citep{mm88,ll91} and therefore
single stars reach the Wolf-Rayet phase at a higher mass when their
metallicity is low; in the Galaxy $M_{\rm WR,min}\approx 25$
M$_\odot$, while in the SMC $M_{\rm WR,min}\approx 45$ M$_\odot$
\citep{mae98,mm00}.  With a metallicity of $\lesssim 0.1$ $Z_\odot$ 
and a mass of only $\sim 25$ M$_\odot$, it appears that the progenitor
of \grb\ was not sufficiently massive to have become a Wolf-Rayet
star.  It is therefore likely that the progenitor required a companion
star in order to eject the hydrogen envelope.

\subsection{\ion{Si}{2}*}

\begin{figure}
\centerline{\psfig{file=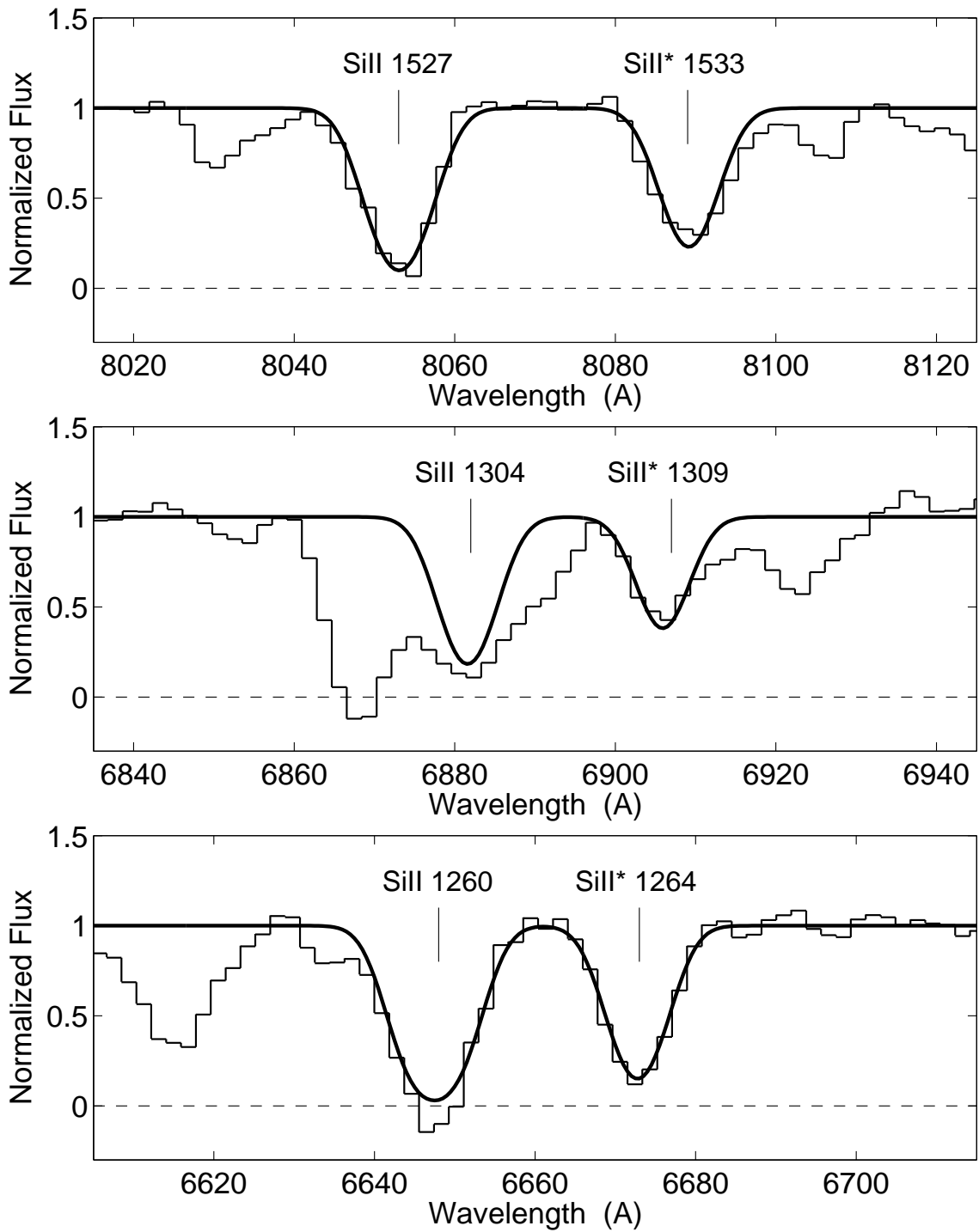,width=6.0in}}
\caption{Absorption lines of \ion{Si}{2} and the fine structure 
\ion{Si}{2}*.   The lines of both species appear to be saturated 
so the ratio of column densities is only roughly determined to be
$\sim 0.2$.  We note that \ion{Si}{2}$\lambda 1304$ and
\ion{Si}{2}*$\lambda 1309$ are located in the atmospheric B band 
and the derived equivalent widths are highly uncertain.
\label{fig:siii}}
\end{figure}

Another interesting feature of the spectrum of \grb\ is the detection
of strong fine-structure absorption from \ion{Si}{2}*.  Since these
transitions have not been convincingly detected in QSO-DLAs (e.g.,
\citealt{hwp05}) and since they are most likely excited by collisions 
with electrons (for $T\lesssim 10^5$ K; \citealt{sv02}), we conclude
that the \ion{Si}{2}* absorber is coincident with the local, high
density environment of the GRB.  In Figure~\ref{fig:siii} we show the
three detected \ion{Si}{2} and \ion{Si}{2}* transitions.  We note that
the \ion{Si}{2}$\lambda 1304$ and \ion{Si}{2}*$\lambda 1309$ pair is
located in the atmospheric B-band, therefore precluding a clear
measurement of the equivalent widths.  In addition, due to the low
spectral resolution, the \ion{Si}{2}$\lambda 1260$ transition is
blended with \ion{S}{2}$\lambda 1259$ and \ion{Fe}{2}$\lambda 1260$,
although given the large oscillator strength of the \ion{Si}{2}
transition, it likely dominates the measured equivalent width.  Our
estimate of the column density is thus based primarily on
\ion{Si}{2}$\lambda 1527$ and on \ion{Si}{2}*$\lambda\lambda 
1264,1533$ (Figure~\ref{fig:siii}).

The absorption lines of both species are saturated and the ratio of
column densities is therefore somewhat uncertain, although it is clear
that the column of \ion{Si}{2}* is fairly large; using the weakest,
unblended line of \ion{Si}{2}* under the assumption of the optically
thin case we find ${\rm log}\,N({\rm SiII^*})\approx 14.7$.  This
value is somewhat higher than those found in two previous GRB spectra:
${\rm log}\,N\approx 14.2$ (GRB\,030323; \citealt{vel+04}) and ${\rm
log}\,N\approx 14.3$ (GRB\,020813; \citealt{sf04}), but the rough
similarity may reflect comparable physical conditions in the local
environments of GRBs.

Using the COG analysis we find that the ratio of column densities is
\begin{equation}
\frac{N_{\rm Si\,II^*}}{N_{\rm Si\,II}}\approx \frac{10^{15.1}}
{10^{15.7}}\approx 0.2.
\end{equation}
With this value, and using the calculations of \citet{sv02}, we
estimate that for a temperature of $10^3$ K (i.e., a neutral medium),
the implied volume density of \ion{H}{1} is $n_{\rm H\,I}\sim 10^2$
cm$^{-3}$ if the electron fraction is $n_e=0.1 n_{\rm H\,I}$.  If the
electron fraction is $n_e<10^{-4}n_{\rm H\,I}$, as may be expected for
a neutral medium in which the electrons primarily come from
low-ionization species with an abundance relative to hydrogen of $\sim
10^{-4}$ \citep{sv02,vel+04}, the density is $n_{\rm H\,I}\sim 10^4$
cm$^{-3}$.  In this framework, the size of the absorbing region is
$\ell\sim 10^{22.05}/10^3\sim 3.5$ pc assuming that the hydrogen is
co-located with the \ion{Si}{2}* absorber.  The inferred mass of the
absorbing region is $M=m_pN(\rm H\,I)\ell^2\sim 10^3$ M$_\odot$.
Naturally, if only a fraction of the \ion{H}{1} column is associated
with the \ion{Si}{2}* absorber, then the derived mass and size are in
fact upper limits.  The inferred compact size of the absorber explains
the non-detection of \ion{Si}{2}* in the QSO-DLAs, since the
probability of intersecting a region of only a few parsec in size is
vanishingly small.

We note that there are two difficulties with this interpretation if we
insist that the compact \ion{Si}{2}* absorber has to be placed around
the GRB.  First, in order to survive the burst of ionizing radiation
from the GRB (which affects the environment on a scale of $\sim 10$
pc) the \ion{Si}{2} absorber has to be shielded.  Second, as discussed
in the previous section, the wind from the progenitor star appears to
be deficient in silicon, and another source for the \ion{Si}{2} may be
required.  These arguments suggest that the \ion{Si}{2} is located in
the vicinity of the GRB, but is probably not directly associated with
the progenitor star itself, and is instead a signature of the dense
environment in the star forming region.

\subsection{Limits on the ${\rm H}_2$ Column Density}
\label{sec:h2}

Absorption lines of molecular hydrogen may be detected in the spectrum
of \grb\ if the burst occurred in a dense molecular cloud.
Unfortunately, at the low resolution of our spectrum it is essentially
impossible to discern ${\rm H}_2$ absorption lines from the Ly$\alpha$
forest and transitions due to metals.  We note that in general the
molecular fraction in DLAs is $\lesssim 10^{-4}$, and only a small
number of DLAs exhibit detectable fractions of ${\rm H}_2$
\citep{lps03}.  The explanation for this trend is a combination of a
low dust content and fairly intense far-UV radiation field, which
combine to reduce the formation rate of ${\rm H}_2$ and increase the
photodissociation rate, respectively \citep{wgp05}.  However, if GRBs
tend to occur in molecular clouds, GRB-DLAs may exhibit preferentially
higher fractions of molecular hydrogen, particularly in light of the
fairly large dust-to-gas ratios in comparison to QSO-DLAs
\citep{sff03}.

On the other hand, as discussed in detail by \citet{dh02}, the
optical/UV radiation from the GRB itself is expected to photoionize
and dissociate ${\rm H}_2$ molecules to a distance of a few parsec.
If the cloud hosting the GRB progenitor is smaller than this size, or
if the burst is located near the edge of the cloud we expect that no
${\rm H}_2$ absorption features will be detected even if the molecular
fraction was initially high.

If the cloud is sufficiently large, the UV radiation field is expected
to also excite ${\rm H}_2$ into vibrational levels which will produce
observable absorption features at rest-wavelengths of $1110-1705$\AA\
\citep{dh02}.  The strongest features are at $\lambda=1254.8$\AA\ and 
$\lambda= 1277.3$\AA.  While we detect an absorption feature
coincident with the 1277.3\AA\ line, we interpret this feature as
\ion{C}{1}$\lambda 1277.2$ whose equivalent width is in good agreement
with other \ion{C}{1} lines.  The non-detection of vibrational ${\rm
H}_2$ absorption, under the assumption that the cloud size is
sufficiently large, places a limit of ${\rm log}\,N({\rm H}_2^*)\lesssim
18$ cm$^{-2}$ \citep{dh02}, and hence $f({\rm H}_2)\lesssim 10^{-4}$.

\subsection{Limits on Ly$\alpha$ Emission}
\label{sec:lyalpha}

We do not detect Ly$\alpha$ in emission with a $3\sigma$ limit of
$F<2.1\times 10^{-17}$ erg cm$^{-2}$ s$^{-1}$.  At the redshift of the
burst this limit translates to an upper limit on the line luminosity
of $L<3.9\times 10^{42}$ erg s$^{-1}$.  Using the conversion of
\citet{ken98} and assuming case B recombination, the limit on the star
formation rate implied by the non-detection of Ly$\alpha$ emission is
${\rm SFR}=9.1\times 10^{-43} L_{\rm Ly\alpha}<3.5$ M$_\odot$
yr$^{-1}$.

The limit on the inferred star formation rate is about a factor of
three higher than the value inferred for GRB\,030323 at $z=3.372$
\citep{vel+04}, but it is significantly lower than the value measured 
for GRB\,021004, $\approx 11$ M$_\odot$ yr$^{-1}$, based on Ly$\alpha$
emission (e.g., \citealt{fgs+05}).  In fact, nearly every GRB host in
which Ly$\alpha$ emission can be detected has been detected
\citep{jbf+05}, while the fraction of LBGs with Ly$\alpha$ equivalent
widths similar to those in GRB hosts is only $\sim 1/3$
\citep{ssp+03}.

We note that the lack of Ly$\alpha$ emission may still accommodate a
star formation rate higher than our formal limit since the Ly$\alpha$
photons are easily destroyed by resonant scattering in the presence of
dust.  Thus, even for a fairly low dust content it is possible that
the star formation rate is in fact higher than the limit provided
above.  A better assessment of the star formation rate requires
near-IR spectroscopy of the host galaxy for detection of H$\alpha$ at
$3.46$ $\mu$m and/or [\ion{O}{2}]$\lambda 3727$ at 1.97 $\mu$m.

\section{Intervening Systems}
\label{sec:inter}

We detect two intervening systems at $z_2=2.265$ and $z_3=1.695$.  The
latter is detected only in \ion{Mg}{2} with a rest-frame equivalent
width of 1.98\AA\ ($\lambda 2796.35$) and 0.94\AA\ ($\lambda
2803.53$).  These values match the expected $2:1$ ratio in the
optically thin case, suggesting that the lines are not saturated.  The
derived column density is about $5\times 10^{13}$ cm$^{-2}$.  A
comparison to the sample of \citet{ss92} reveals that $\sim 10\%$ of
quasar sight lines reveal systems with similarly large equivalent
widths.  However, unlike the observed doublet ratio of about two
measured here, the range of doublet ratios for the quasar \ion{Mg}{2}
systems with ${\rm EW}\gtrsim 2$ \AA\ is typically $1-1.3$, indicative
of strong saturation.  We finally note that for a magnesium abundance
of about $10\%$ the solar value, the inferred column density in the
$z_3$ system suggests an associated neutral hydrogen column density of
${\rm log}\,N({\rm HI})\sim 18.5$ cm$^{-2}$.

The $z_2$ redshift system exhibits a wider range of absorption
features, but unfortunately, some of these lines are blended with
features from the host galaxy.  Still, several \ion{Fe}{2} lines, as
well as the \ion{Mg}{2} doublet, and a single \ion{Mn}{2} line are not
blended, allowing us to estimate the column density of some metals in
the $z_2$ system.  A COG fit to the three \ion{Fe}{2} lines indicates
$b\approx 60$ km s$^{-1}$ and ${\rm log}\,N({\rm FeII})\approx 15.0$.
The \ion{Mg}{2} ratio of about 0.85 indicates strong saturation, and a
lower limit on the column density of ${\rm log}\,N({\rm MgII})\approx
14.6$, assuming the same Doppler parameter as for the \ion{Fe}{2}
lines.  Finally, assuming that the single \ion{Mn}{2} line is
optically thin we estimate ${\rm log}\,N({\rm MnII})\approx 13.7$.
Thus, the relative abundance is ${\rm [Fe/Mn]}\gtrsim -0.7$, while in
DLAs it is typically $\sim 0$.  This suggests that we may be
underestimating the \ion{Fe}{2} by about 0.7 dex.

We also note that both iron and manganese are strongly depleted in the
Galaxy, while elements such as zinc are relatively undepleted
\citep{ss96}.  In our case, the \ion{Zn}{2}$\lambda 2062.66$ line is
blended with \ion{C}{1} from the host galaxy system.  However, using
the measured equivalent width as an upper limit and assuming the
optically thin case, we find ${\rm log}\,N({\rm ZnII})<13.4$.  Thus,
${\rm [Mn/Zn]}>-0.5$ and ${\rm [Fe/Zn]}>-1.2$ are in agreement with
the typical values measured for DLAs of about $-0.5$ and $-0.4$,
respectively, particularly if the abundance of \ion{Fe}{2} is in fact
underestimated as suggested above.

\section{Discussion and Conclusions}
\label{sec:disc}

\begin{figure}
\centerline{\psfig{file=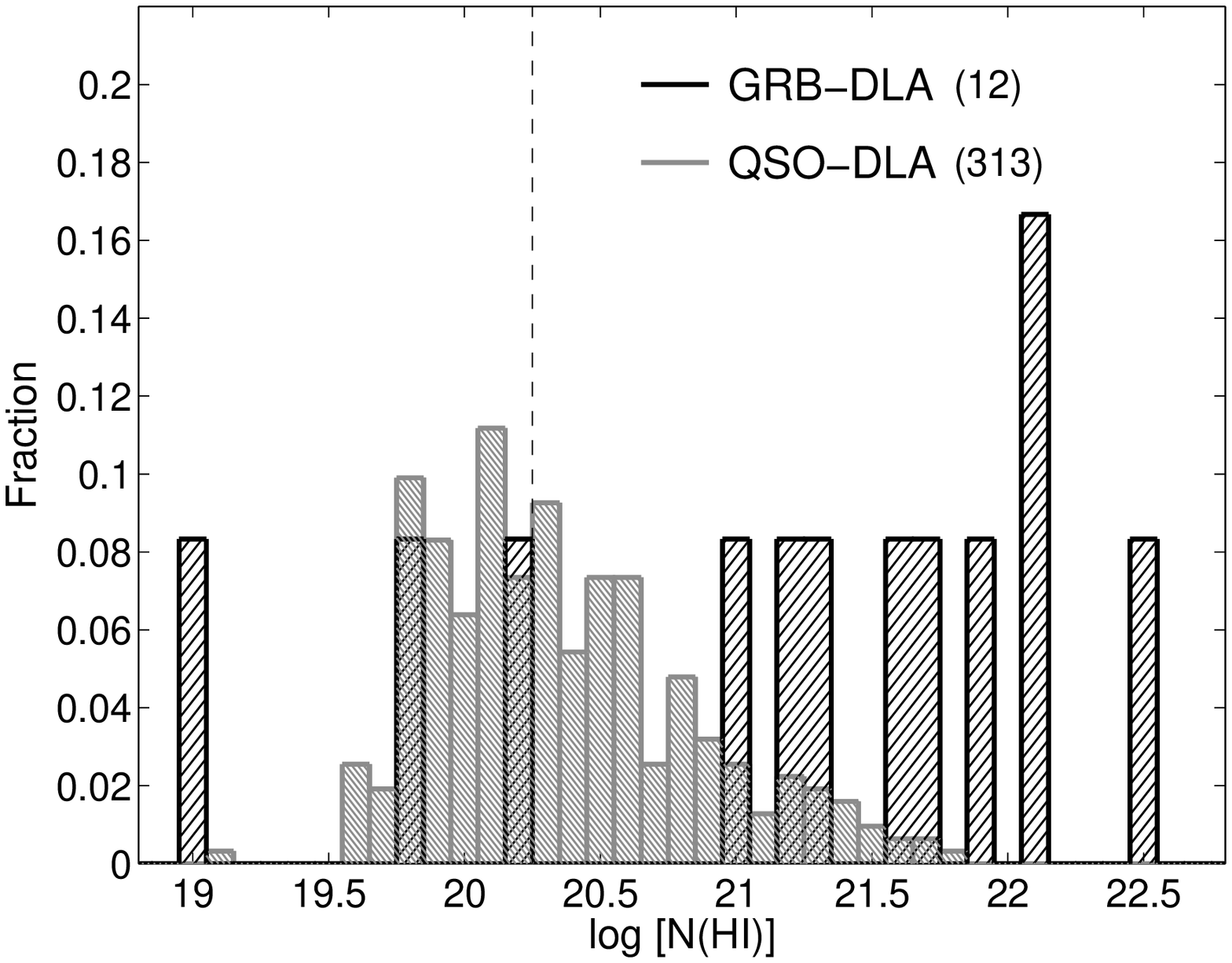,width=6.5in}}
\caption{Fractional distribution of neutral hydrogen column 
densities from quasar \citep{cwm+02} and GRB observations
\citep{jhf+04,vel+04,cpb+05,pfc+05,pbw+05,sve+05,vsf+05,wfl+05}.  
The GRB-DLAs exhibit higher column densities with about a third
exceeding the values measured in any QSO-DLA.
\label{fig:dla}}
\end{figure}

The absorption spectrum of GRB\,050505 exhibits a DLA with a column
density of ${\rm log}\,N({\rm HI})\approx 22.05$ associated with the
host galaxy.  The measured column is higher than that of any QSO-DLA
studied to date, but is similar to those measured for several
GRB-DLAs, albeit at the highest redshift to
date\footnotemark\footnotetext{An absorption spectrum of GRB\,050904
at $z=6.29$ has been obtained by \citet{kyk+05}, but the column
density and metallicity have not been published so far.}.  In
Figure~\ref{fig:dla} we provide a summary of the column densities
measured from both quasars and GRBs to date.  As already noted in
\citet{vel+04}, these distributions differ significantly, presumably
as a result of the smaller impact parameters in the case of GRB-DLAs
and the possibility that they probe individual dense star forming
regions.  The metallicity we infer for \grb, $Z\approx 0.06$
$Z_\odot$, is within the range of values measured for QSO-DLAs at a
similar redshift \citep{pgw+03}, but is a factor of several higher
than the mean metallicity of about $0.01$ $Z_\odot$.

\begin{figure}
\centerline{\psfig{file=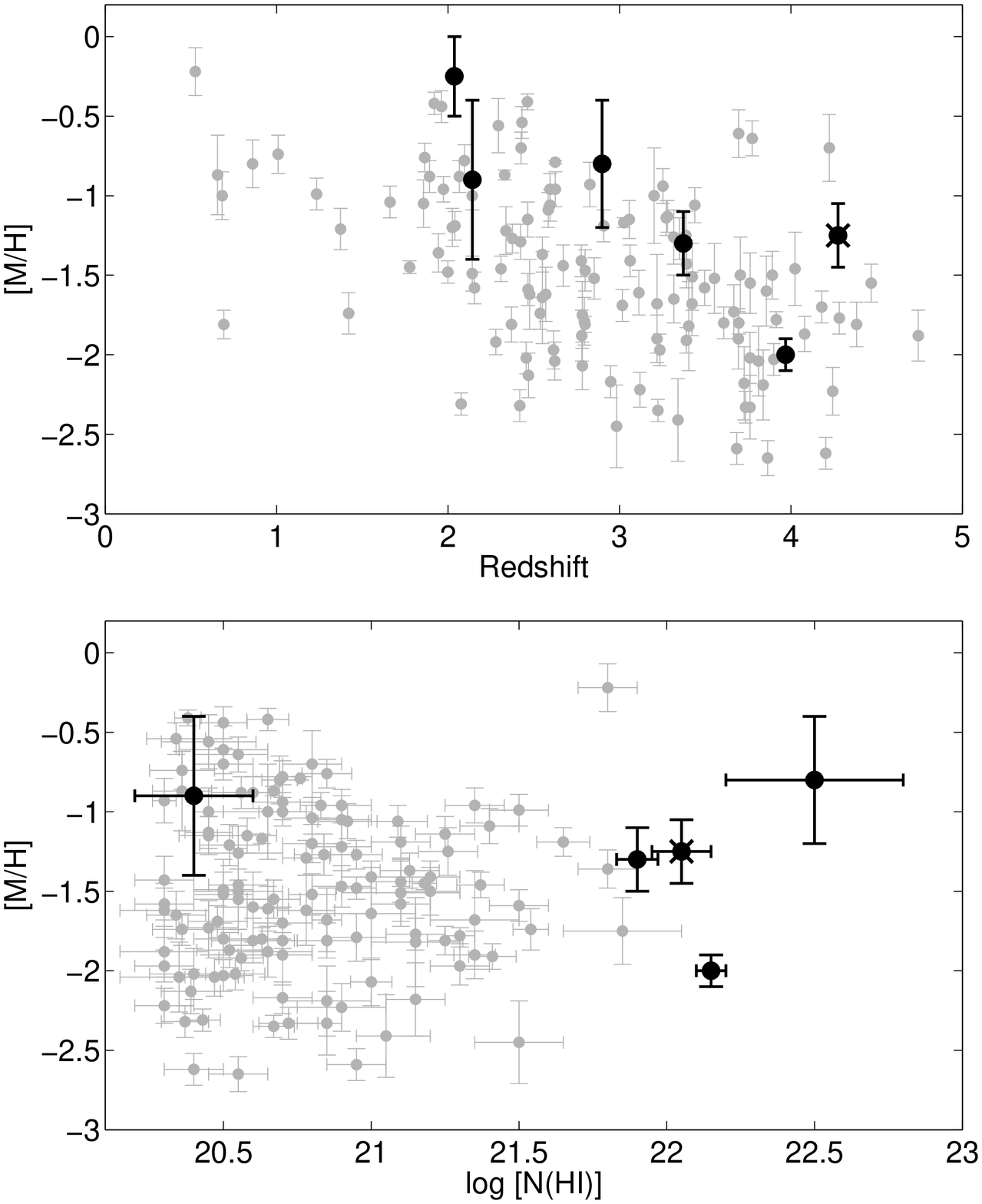,width=6.0in}}
\caption{{\it Top:} Metallicity as a function of redshift for 
QSO-DLAs (gray; \citealt{pgw+03}) and GRB-DLAs (black;
\citealt{sff03,vel+04,cpb+05,sve+05,vsf+05,wfl+05});
\grb\ is marked with a cross.  {\it Bottom:} Metallicity as a 
function of neutral hydrogen column density.  The trend of increased
metallicity with lower redshift, noted for QSO-DLAs by \citet{pgw+03},
is also apparent in the GRB-DLA sample, albeit at an overall higher
metallicity.  This suggests that GRBs select more metal-enriched
systems, but it is not clear at the present if this is related to the
higher column densities.
\label{fig:metal_z}} 
\end{figure}

This, along with the metallicities inferred for five additional
GRB-DLAs, is shown in Figure~\ref{fig:metal_z}.  In the top panel we
plot the metallicity as determined from various indicators as a
function of redshift in comparison to QSO-DLAs.  For the QSO sample,
\citet{pgw+03} noted an apparent trend of increased metallicity with
lower redshift.  A similar trend appears in the admittedly small
GRB-DLA sample which spans from $z\approx 2$ to 4.3.  However, it is
also clear that the GRB-DLAs have systematically higher metallicities
compared to QSO-DLAs.  In the bottom panel of Figure~\ref{fig:metal_z}
we plot the metallicity as a function of hydrogen column density.  No
clear trend is evident, perhaps suggesting that the higher
metallicities in GRB-DLAs may not be simply a function of the column
density, but may reflect a difference in the star formation
properties, rates, or history of GRB-selected host galaxies compared
to QSO-DLAs.

We highlight two aspects of the spectrum, which provide insight into
scales and conditions that are not typically probed in quasar sight
lines: A \ion{C}{4} outflow that is most likely associated with the
wind of the progenitor star, and strong absorption from fine-structure
\ion{Si}{2}* that is associated with a dense region in the vicinity of
the burst.  The latter suggests that the GRB progenitor lived in an
environment typical of a molecular cloud.  The \ion{C}{4} outflow,
taken in association with the lack of corresponding \ion{Si}{4}
absorption suggests that the progenitor star had a relatively low mass
and metallicity, $M\sim 25$ M$_\odot$ and $Z\lesssim 0.1$ $Z_\odot$.
The combination of these properties suggests that the progenitor was
part of a binary system, with the companion responsible for stripping
the hydrogen envelope.

The continued detection of high-velocity features in GRB absorption
spectra may thus hold the only key to a detailed understanding of the
distribution of progenitor properties (mass, metallicity, binarity),
as well as the ability to probe the properties of individual molecular
clouds at high redshift.  Coupled with information on the galactic
scale properties of the ISM, GRBs can probe all the scales relevant to
our understanding of the star formation process and its relation to
metal production, as well as the relation between DLAs and star
forming galaxies.

\acknowledgements
We thank Michael Rauch for helpful discussions and comments, and
Allard-Jan van Marle for information on Wolf-Rayet winds.  EB is
supported is supported by NASA through Hubble Fellowship grant
HST-01171.01 awarded by the Space Telescope Science Institute, which
is operated by AURA, Inc., for NASA under contract NAS 5-26555.


\begin{thebibliography}{}
 
\bibitem[\protect\citeauthoryear{{Abbott}}{{Abbott}}{1978}]{abb78}
{Abbott}, D.~C. 1978, \apj, 225, 893
 
\bibitem[\protect\citeauthoryear{{Adelberger} et~al.}{{Adelberger}
  et~al.}{2003}]{ass+03}
{Adelberger}, K.~L., {Steidel}, C.~C., {Shapley}, A.~E.,  \& {Pettini}, M.
  2003, \apj, 584, 45
 
\bibitem[\protect\citeauthoryear{{Asplund}, {Grevesse}, \& {Sauval}}{{Asplund}
  et~al.}{2005}]{ags05}
{Asplund}, M., {Grevesse}, N.,  \& {Sauval}, A.~J. 2005, in ASP Conf. Ser. 336:
  Cosmic Abundances as Records of Stellar Evolution and Nucleosynthesis, 25
 
\bibitem[\protect\citeauthoryear{{Berger} et~al.}{{Berger}
  et~al.}{2005}]{bkf+05}
{Berger}, E., et~al. 2005, ArXiv Astrophysics e-prints
 
\bibitem[\protect\citeauthoryear{{Berger}, {Kulkarni}, \& {Frail}}{{Berger}
  et~al.}{2003}]{bkf03}
{Berger}, E., {Kulkarni}, S.~R.,  \& {Frail}, D.~A. 2003, \apj, 590, 379
                                                                                          
\bibitem[\protect\citeauthoryear{{Bloom}, {Frail}, \& {Kulkarni}}{{Bloom}
  et~al.}{2003}]{bfk03}
{Bloom}, J.~S., {Frail}, D.~A.,  \& {Kulkarni}, S.~R. 2003, \apj, 594, 674
 
\bibitem[\protect\citeauthoryear{{Bloom}, {Kulkarni}, \& {Djorgovski}}{{Bloom}
  et~al.}{2002}]{bkd02}
{Bloom}, J.~S., {Kulkarni}, S.~R.,  \& {Djorgovski}, S.~G. 2002, \aj, 123, 1111
 
\bibitem[\protect\citeauthoryear{{Bunker} et~al.}{{Bunker}
  et~al.}{1999}]{bwc+99}
{Bunker}, A.~J., {Warren}, S.~J., {Clements}, D.~L., {Williger}, G.~M.,  \&
  {Hewett}, P.~C. 1999, \mnras, 309, 875
 
\bibitem[\protect\citeauthoryear{{Castro} et~al.}{{Castro}
  et~al.}{2003}]{cgh+03}
{Castro}, S., {Galama}, T.~J., {Harrison}, F.~A., {Holtzman}, J.~A., {Bloom},
  J.~S., {Djorgovski}, S.~G.,  \& {Kulkarni}, S.~R. 2003, \apj, 586, 128
 
\bibitem[\protect\citeauthoryear{{Chen} \& {Lanzetta}}{{Chen} \&
  {Lanzetta}}{2003}]{cl03}
{Chen}, H.-W.,  \& {Lanzetta}, K.~M. 2003, \apj, 597, 706
                                                                                          
\bibitem[\protect\citeauthoryear{{Chen} et~al.}{{Chen} et~al.}{2005}]{cpb+05}
{Chen}, H.-W., {Prochaska}, J.~X., {Bloom}, J.~S.,  \& {Thompson}, I.~B. 2005,
  ArXiv Astrophysics e-prints
 
\bibitem[\protect\citeauthoryear{{Christensen}, {Hjorth}, \&
  {Gorosabel}}{{Christensen} et~al.}{2004}]{chg04}
{Christensen}, L., {Hjorth}, J.,  \& {Gorosabel}, J. 2004, \aap, 425, 913
 
\bibitem[\protect\citeauthoryear{{Cristiani} et~al.}{{Cristiani}
  et~al.}{1993}]{cgb+93}
{Cristiani}, S., {Giallongo}, E., {Buson}, L.~M., {Gouiffes}, C.,  \& {La
  Franca}, F. 1993, \aap, 268, 86
 
\bibitem[\protect\citeauthoryear{{Curran} et~al.}{{Curran}
  et~al.}{2002}]{cwm+02}
{Curran}, S.~J., {Webb}, J.~K., {Murphy}, M.~T., {Bandiera}, R., {Corbelli},
  E.,  \& {Flambaum}, V.~V. 2002, Publications of the Astronomical Society of
  Australia, 19, 455
 
\bibitem[\protect\citeauthoryear{{Draine} \& {Hao}}{{Draine} \&
  {Hao}}{2002}]{dh02}
{Draine}, B.~T.,  \& {Hao}, L. 2002, \apj, 569, 780
 
\bibitem[\protect\citeauthoryear{{Fiore} et~al.}{{Fiore} et~al.}{2005}]{fdl+05}
{Fiore}, F., et~al. 2005, \apj, 624, 853
 
\bibitem[\protect\citeauthoryear{{F{\"o}rster Schreiber} et~al.}{{F{\"o}rster
  Schreiber} et~al.}{2004}]{fdf+04}
{F{\"o}rster Schreiber}, N.~M., et~al. 2004, \apj, 616, 40
 
\bibitem[\protect\citeauthoryear{{Frye}, {Broadhurst}, \&
  {Ben{\'{\i}}tez}}{{Frye} et~al.}{2002}]{fbb02}
{Frye}, B., {Broadhurst}, T.,  \& {Ben{\'{\i}}tez}, N. 2002, \apj, 568, 558
 
\bibitem[\protect\citeauthoryear{{Fynbo} et~al.}{{Fynbo} et~al.}{2005}]{fgs+05}
{Fynbo}, J.~P.~U., et~al. 2005, \apj, 633, 317
 
\bibitem[\protect\citeauthoryear{{Haislip} et~al.}{{Haislip}
  et~al.}{2005}]{hnr+05}
{Haislip}, J., et~al. 2005, ArXiv Astrophysics e-prints
 
\bibitem[\protect\citeauthoryear{{Heckman} et~al.}{{Heckman}
  et~al.}{2000}]{hls+00}
{Heckman}, T.~M., {Lehnert}, M.~D., {Strickland}, D.~K.,  \& {Armus}, L. 2000,
  \apjs, 129, 493
 
\bibitem[\protect\citeauthoryear{{Howk}, {Wolfe}, \& {Prochaska}}{{Howk}
  et~al.}{2005}]{hwp05}
{Howk}, J.~C., {Wolfe}, A.~M.,  \& {Prochaska}, J.~X. 2005, \apjl, 622, L81
 
\bibitem[\protect\citeauthoryear{{Hullinger} et~al.}{{Hullinger}
  et~al.}{2005}]{gcn3364}
{Hullinger}, D., et~al. 2005, GRB Circular Network, 3364, 1
 
\bibitem[\protect\citeauthoryear{{Jakobsson} et~al.}{{Jakobsson}
  et~al.}{2005a}]{jbf+05}
{Jakobsson}, P., et~al. 2005a, \mnras, 362, 245
 
\bibitem[\protect\citeauthoryear{{Jakobsson} et~al.}{{Jakobsson}
  et~al.}{2004}]{jhf+04}
{Jakobsson}, P., et~al. 2004, \aap, 427, 785
 
\bibitem[\protect\citeauthoryear{{Jakobsson} et~al.}{{Jakobsson}
  et~al.}{2005b}]{jlf+05}
{Jakobsson}, P., et~al. 2005b, ArXiv Astrophysics e-prints
                                                                                          
\bibitem[\protect\citeauthoryear{{Kawai} et~al.}{{Kawai} et~al.}{2005}]{kyk+05}
{Kawai}, N., {Yamada}, T., {Kosugi}, G., {Hattori}, T.,  \& {Aoki}, K. 2005,
  GRB Circular Network, 3937, 1
 
\bibitem[\protect\citeauthoryear{{Kelson}}{{Kelson}}{2003}]{kel03}
{Kelson}, D.~D. 2003, \pasp, 115, 688
 
\bibitem[\protect\citeauthoryear{{Kennea} et~al.}{{Kennea}
  et~al.}{2005}]{gcn3365}
{Kennea}, J.~A., {Burrows}, D.~N., {Hurkett}, C.~P., {Osbourne}, J.~P.,  \&
  {Gehrels}, N. 2005, GRB Circular Network, 3365, 1
 
\bibitem[\protect\citeauthoryear{{Kennicutt}}{{Kennicutt}}{1998}]{ken98}
{Kennicutt}, R.~C. 1998, \araa, 36, 189
 
\bibitem[\protect\citeauthoryear{{Ledoux}, {Petitjean}, \& {Srianand}}{{Ledoux}
  et~al.}{2003}]{lps03}
{Ledoux}, C., {Petitjean}, P.,  \& {Srianand}, R. 2003, \mnras, 346, 209
 
\bibitem[\protect\citeauthoryear{{Leitherer} \& {Lamers}}{{Leitherer} \&
  {Lamers}}{1991}]{ll91}
{Leitherer}, C.,  \& {Lamers}, H.~J.~G.~L. 1991, \apj, 373, 89
 
\bibitem[\protect\citeauthoryear{{Lopez} \& {Ellison}}{{Lopez} \&
  {Ellison}}{2003}]{le03}
{Lopez}, S.,  \& {Ellison}, S.~L. 2003, \aap, 403, 573
 
\bibitem[\protect\citeauthoryear{{Maeder}}{{Maeder}}{1998}]{mae98}
{Maeder}, A. 1998, in IAU Symp. 189: Fundamental Stellar Properties, 313
 
\bibitem[\protect\citeauthoryear{{Maeder} \& {Meynet}}{{Maeder} \&
  {Meynet}}{1988}]{mm88}
{Maeder}, A.,  \& {Meynet}, G. 1988, \aaps, 76, 411
 
\bibitem[\protect\citeauthoryear{{Maeder} \& {Meynet}}{{Maeder} \&
  {Meynet}}{2000}]{mm00}
{Maeder}, A.,  \& {Meynet}, G. 2000, \araa, 38, 143
 
\bibitem[\protect\citeauthoryear{{Mirabal} et~al.}{{Mirabal}
  et~al.}{2003}]{mhc+03}
{Mirabal}, N., et~al. 2003, \apj, 595, 935
 
\bibitem[\protect\citeauthoryear{{M{\"o}ller} et~al.}{{M{\"o}ller}
  et~al.}{2002}]{mfh+02}
{M{\"o}ller}, P., et~al. 2002, \aap, 396, L21
 
\bibitem[\protect\citeauthoryear{{Nousek} et~al.}{{Nousek}
  et~al.}{2005}]{nkg+05}
{Nousek}, J.~A., et~al. 2005, ArXiv Astrophysics e-prints
 
\bibitem[\protect\citeauthoryear{{Perna} \& {Loeb}}{{Perna} \&
  {Loeb}}{1998}]{pl98}
{Perna}, R.,  \& {Loeb}, A. 1998, \apj, 501, 467
 
\bibitem[\protect\citeauthoryear{{Pichon} et~al.}{{Pichon}
  et~al.}{2003}]{psa+03}
{Pichon}, C., {Scannapieco}, E., {Aracil}, B., {Petitjean}, P., {Aubert}, D.,
  {Bergeron}, J.,  \& {Colombi}, S. 2003, \apjl, 597, L97
 
\bibitem[\protect\citeauthoryear{{Porciani} \& {Madau}}{{Porciani} \&
  {Madau}}{2005}]{pm05}
{Porciani}, C.,  \& {Madau}, P. 2005, \apjl, 625, L43
 
\bibitem[\protect\citeauthoryear{{Prochaska} et~al.}{{Prochaska}
  et~al.}{2005a}]{pbw+05}
{Prochaska}, J.~X., {Bloom}, J.~S., {Wright}, J.~T., {Butler}, R.~P., {Chen},
  H.~W., {Vogt}, S.~S.,  \& {Marcy}, G.~W. 2005a, GRB Circular Network, 3833, 1
 
\bibitem[\protect\citeauthoryear{{Prochaska} et~al.}{{Prochaska}
  et~al.}{2005b}]{pfc+05}
{Prochaska}, J.~X., {Foley}, R.~J., {Chen}, H.-W., {Bloom}, J.~S., {Hurley},
  K., {Cooper}, M., {Guhathakurta}, R.,  \& {Li}, W. 2005b, GRB Circular
  Network, 3971, 1
 
\bibitem[\protect\citeauthoryear{{Prochaska} et~al.}{{Prochaska}
  et~al.}{2003}]{pgw+03}
{Prochaska}, J.~X., {Gawiser}, E., {Wolfe}, A.~M., {Cooke}, J.,  \& {Gelino},
  D. 2003, \apjs, 147, 227
 
\bibitem[\protect\citeauthoryear{{Rauch} et~al.}{{Rauch} et~al.}{1996}]{rsw+96}
{Rauch}, M., {Sargent}, W.~L.~W., {Womble}, D.~S.,  \& {Barlow}, T.~A. 1996,
  \apjl, 467, L5
 
\bibitem[\protect\citeauthoryear{{Rol} et~al.}{{Rol} et~al.}{2005}]{gcn3372}
{Rol}, E., {Tanvir}, N., {Levan}, A., {Adamson}, A., {Fuhrman}, L., {Priddey},
  R.,  \& {Chapman}, R. 2005, GRB Circular Network, 3372, 1
 
\bibitem[\protect\citeauthoryear{{Rosen} et~al.}{{Rosen}
  et~al.}{2005}]{gcn3371}
{Rosen}, S., {Hurkett}, C., {Holland}, S., {Roming}, P., {Blustin}, A.,
  {Gehrels}, N., {Mason}, K.,  \& {Nousek}, J. 2005, GRB Circular Network,
  3371, 1
 
\bibitem[\protect\citeauthoryear{{Savage} \& {Sembach}}{{Savage} \&
  {Sembach}}{1996}]{ss96}
{Savage}, B.~D.,  \& {Sembach}, K.~R. 1996, \araa, 34, 279
 
\bibitem[\protect\citeauthoryear{{Savaglio} \& {Fall}}{{Savaglio} \&
  {Fall}}{2004}]{sf04}
{Savaglio}, S.,  \& {Fall}, S.~M. 2004, \apj, 614, 293
 
\bibitem[\protect\citeauthoryear{{Savaglio}, {Fall}, \& {Fiore}}{{Savaglio}
  et~al.}{2003}]{sff03}
{Savaglio}, S., {Fall}, S.~M.,  \& {Fiore}, F. 2003, \apj, 585, 638
 
\bibitem[\protect\citeauthoryear{{Schaefer} et~al.}{{Schaefer}
  et~al.}{2003}]{sgh+03}
{Schaefer}, B.~E., et~al. 2003, \apj, 588, 387
 
\bibitem[\protect\citeauthoryear{{Schlegel}, {Finkbeiner}, \&
  {Davis}}{{Schlegel} et~al.}{1998}]{sfd98}
{Schlegel}, D.~J., {Finkbeiner}, D.~P.,  \& {Davis}, M. 1998, \apj, 500, 525
 
\bibitem[\protect\citeauthoryear{{Shapley} et~al.}{{Shapley}
  et~al.}{2004}]{sep+04}
{Shapley}, A.~E., {Erb}, D.~K., {Pettini}, M., {Steidel}, C.~C.,  \&
  {Adelberger}, K.~L. 2004, \apj, 612, 108
 
\bibitem[\protect\citeauthoryear{{Shapley} et~al.}{{Shapley}
  et~al.}{2001}]{ssa+01}
{Shapley}, A.~E., {Steidel}, C.~C., {Adelberger}, K.~L., {Dickinson}, M.,
  {Giavalisco}, M.,  \& {Pettini}, M. 2001, \apj, 562, 95
 
\bibitem[\protect\citeauthoryear{{Shapley} et~al.}{{Shapley}
  et~al.}{2003}]{ssp+03}
{Shapley}, A.~E., {Steidel}, C.~C., {Pettini}, M.,  \& {Adelberger}, K.~L.
  2003, \apj, 588, 65
 
\bibitem[\protect\citeauthoryear{{Silva} \& {Viegas}}{{Silva} \&
  {Viegas}}{2002}]{sv02}
{Silva}, A.~I.,  \& {Viegas}, S.~M. 2002, \mnras, 329, 135
                                                                                          
\bibitem[\protect\citeauthoryear{{Songaila}}{{Songaila}}{2005}]{son05}
{Songaila}, A. 2005, ArXiv Astrophysics e-prints
 
\bibitem[\protect\citeauthoryear{{Spitzer}}{{Spitzer}}{1978}]{spi78}
{Spitzer}, L. 1978, {Physical processes in the interstellar medium} (New York
  Wiley-Interscience, 1978.~333 p.)
 
\bibitem[\protect\citeauthoryear{{Starling} et~al.}{{Starling}
  et~al.}{2005a}]{sve+05}
{Starling}, R.~L.~C., et~al. 2005a, \aap, 442, L21
 
\bibitem[\protect\citeauthoryear{{Starling} et~al.}{{Starling}
  et~al.}{2005b}]{swh+05}
{Starling}, R.~L.~C., {Wijers}, R.~A.~M.~J., {Hughes}, M.~A., {Tanvir}, N.~R.,
  {Vreeswijk}, P.~M., {Rol}, E.,  \& {Salamanca}, I. 2005b, \mnras, 360, 305
 
\bibitem[\protect\citeauthoryear{{Steidel} et~al.}{{Steidel}
  et~al.}{2003}]{sas+03}
{Steidel}, C.~C., {Adelberger}, K.~L., {Shapley}, A.~E., {Pettini}, M.,
  {Dickinson}, M.,  \& {Giavalisco}, M. 2003, \apj, 592, 728
 
\bibitem[\protect\citeauthoryear{{Steidel} \& {Sargent}}{{Steidel} \&
  {Sargent}}{1992}]{ss92}
{Steidel}, C.~C.,  \& {Sargent}, W.~L.~W. 1992, \apjs, 80, 1
 
\bibitem[\protect\citeauthoryear{{Swinbank} et~al.}{{Swinbank}
  et~al.}{2004}]{ssc+04}
{Swinbank}, A.~M., {Smail}, I., {Chapman}, S.~C., {Blain}, A.~W., {Ivison},
  R.~J.,  \& {Keel}, W.~C. 2004, \apj, 617, 64
 
\bibitem[\protect\citeauthoryear{{van Marle}, {Langer}, \&
  {Garcia-Segura}}{{van Marle} et~al.}{2005}]{mlg05}
{van Marle}, A.-J., {Langer}, N.,  \& {Garcia-Segura}, G. 2005, ArXiv
  Astrophysics e-prints
 
\bibitem[\protect\citeauthoryear{{Vreeswijk} et~al.}{{Vreeswijk}
  et~al.}{2004}]{vel+04}
{Vreeswijk}, P.~M., et~al. 2004, \aap, 419, 927
 
\bibitem[\protect\citeauthoryear{{Vreeswijk} et~al.}{{Vreeswijk}
  et~al.}{2005}]{vsf+05}
{Vreeswijk}, P.~M., et~al. 2005, ArXiv Astrophysics e-prints
 
\bibitem[\protect\citeauthoryear{{Watson} et~al.}{{Watson}
  et~al.}{2005}]{wfl+05}
{Watson}, D., et~al. 2005, ArXiv Astrophysics e-prints
 
\bibitem[\protect\citeauthoryear{{Wolfe}, {Gawiser}, \& {Prochaska}}{{Wolfe}
  et~al.}{2005}]{wgp05}
{Wolfe}, A.~M., {Gawiser}, E.,  \& {Prochaska}, J.~X. 2005, \araa, 43, 861
 
\end{thebibliography}

\clearpage
\begin{deluxetable}{llllll}
\tablecolumns{6}
\tabcolsep0.2in\footnotesize
\tablewidth{0pc}
\tablecaption{Line Identification
\label{tab:lines}}
\tablehead {
\colhead {$\lambda_{\rm obs}$}         &
\colhead {Line}                        &
\colhead {$f_{ij}$}                    &
\colhead {$z$}                         &
\colhead {$W_0$}                       &
\colhead {${\rm log}\,N$}                \\
\colhead {(\AA)}                       &
\colhead {(\AA)}                       &
\colhead {}                            &
\colhead {}                            &
\colhead {(\AA)}                       &
\colhead {(cm$^{-2}$)}                       
}
\startdata
6597.80 & \ion{S}{2}  1250.584         & $5.453\times 10^{-3}$ & 4.2758 & 0.72 &  16.0 \\
6614.47 & \ion{S}{2}  1253.811         & $1.088\times 10^{-2}$ & 4.2755 & 1.27 &  15.9 \\
        & \ion{Zn}{2} 2026.136         & $0.489$               & 2.2646 & \nod & \nod  \\
        & \ion{Cr}{2} 2026.269         & $4.710\times 10^{-3}$ & 2.2644 & \nod & \nod  \\
        & \ion{Mg}{1} 2026.477         & $0.112$               & 2.2640 & \nod &  \nod \\
6646.34 & \ion{S}{2}  1259.519         & $1.624\times 10^{-2}$ & 4.2769 & 2.60 &  16.1 \\
        & \ion{Si}{2} 1260.422         & $1.007$               & 4.2731 & \nod &  14.3 \\
        & \ion{Fe}{2} 1260.533         & $2.500\times 10^{-2}$ & 4.2726 & \nod &  15.9 \\
6671.86 & \ion{Si}{2}$^*$ 1264.738     & $0.903$               & 4.2753 & 1.65 &  14.1 \\
6736.97 & \ion{C}{1}  1277.245         & $9.665\times 10^{-2}$ & 4.2747 & 0.51 &  14.6 \\
        & \ion{Cr}{2} 2062.234         & $7.800\times 10^{-2}$ & 2.2668 & \nod & \nod  \\
        & \ion{Zn}{2} 2062.664         & $0.256$               & 2.2661 & \nod & \nod  \\
6866.63$^a$ & \ion{O}{1}  1302.168     & $4.887\times 10^{-2}$ & 4.2732 & 1.51 &  15.3 \\
6880.58$^a$ & \ion{Si}{2} 1304.370     & $9.400\times 10^{-2}$ & 4.2750 & 3.25 &  15.4 \\
6905.64$^a$ & \ion{Si}{2}$^*$ 1309.275 & $8.600\times 10^{-2}$ & 4.2744 & 1.07 &  14.9 \\
6921.57$^a$ & ?                        & \nod                  & \nod   & \nod & \nod  \\
6949.48 & \ion{Ni}{2} 1317.217         & $7.786\times 10^{-2}$ & 4.2759 & 0.22 &  14.3 \\
7009.27 & \ion{C}{1}  1328.833         & $5.804\times 10^{-2}$ & 4.2748 & 0.37 &  14.6 \\
7040.95 & \ion{C}{2}  1334.532         & $0.128$               & 4.2760 & 3.37 &  15.2 \\
        & \ion{C}{2}$^*$ 1335.708      & $0.115$               & 4.2713 & \nod &  15.3 \\
7350.47 & \ion{Si}{4} 1393.755         & $0.528$               & 4.2739 & 1.77 &  14.3 \\
        & \ion{Fe}{2} 2249.877         & $1.821\times 10^{-3}$ & 2.2671 & \nod & \nod  \\
7399.07 & \ion{Si}{4} 1402.770         & $0.262$               & 4.2746 & 1.60 &  14.5 \\
7536.67 & \ion{Mg}{2} 2796.352         & $0.612$               & 1.6952 & 1.98 &  13.7 \\
7554.21 & \ion{Mg}{2} 2803.531         & $0.305$               & 1.6945 & 0.94 &  13.7 \\
7604.14$^b$ & $\oplus$                 & \nod                  & \nod   & \nod & \nod  \\
7625.86$^b$ & $\oplus$                 & \nod                  & \nod   & \nod & \nod  \\
7639.15$^b$ & ?                        & \nod                  & \nod   & \nod & \nod  \\
7654.06$^b$ & \ion{Fe}{2} 2344.214     & $0.114$               & 2.2651 & 2.29 &  14.6 \\
7676.48$^b$ & \ion{Ni}{2} 1454.842     & $3.230\times 10^{-2}$ & 4.2765 & 1.36 &  15.4 \\
7740.67 & \ion{Ni}{2} 1467.259         & $6.300\times 10^{-3}$ & 4.2756 & 0.25 &  15.3 \\    
        & \ion{Ni}{2} 1467.756         & $9.900\times 10^{-3}$ & 4.2738 & \nod &  15.1 \\    
7752.13 & \ion{Fe}{2} 2374.461         & $3.130\times 10^{-2}$ & 2.2648 & 0.87 &  14.7 \\
7779.65 & \ion{Fe}{2} 2382.765         & $0.320$               & 2.2650 & 1.52 &  14.0 \\
8030.42 & ?                            & \nod                  & \nod   & \nod & \nod  \\
8051.97 & \ion{Si}{2} 1526.707         & $0.127$               & 4.2741 & 1.81 &  14.8 \\
8088.55 & \ion{Si}{2}$^*$ 1533.432     & $0.132$               & 4.2748 & 1.58 &  14.8 \\
8104.98 & ?                            & \nod                  & \nod   & \nod & \nod  \\
8138.73 & \ion{C}{4} 1548.195          & $0.191$               & 4.2569 & 3.23 &  14.9 \\
8151.23 & \ion{C}{4} 1550.770          & $9.522\times 10^{-2}$ & 4.2562 & 2.99 &  15.2 \\
8164.63 & \ion{C}{4} 1548.195          & $0.191$               & 4.2736 & 2.69 &  14.8 \\
8177.95 & \ion{C}{4} 1550.770          & $9.522\times 10^{-2}$ & 4.2735 & 2.64 &  15.1 \\
8221.83 & \ion{C}{1} 1560.309          & $8.041\times 10^{-2}$ & 4.2694 & 1.39 &  14.9 \\
        & \ion{C}{1}$^*$ 1560.682      & $6.030\times 10^{-2}$ & 4.2681 & \nod &  15.0 \\
        & \ion{C}{1}$^*$ 1560.709      & $2.010\times 10^{-2}$ & 4.2680 & \nod &  15.5 \\
8411.32 & \ion{Mn}{2} 2576.877         & $0.351$               & 2.2642 & 0.95 &  13.7 \\
8446.68 & \ion{Fe}{2} 2586.650         & $6.910\times 10^{-2}$ & 2.2655 & 1.70 &  14.6 \\
8487.59 & \ion{Fe}{2} 1608.451         & $5.800\times 10^{-2}$ & 4.2769 & 2.73 &  15.3 \\
        & \ion{Fe}{2} 2600.173         & $0.239$               & 2.2642 & \nod & \nod  \\
8505.42 & \ion{Fe}{2} 1611.200         & $1.360\times 10^{-3}$ & 4.2789 & 0.54 &  16.2 \\
        & \ion{Mn}{2} 2606.462         & $0.193$               & 2.2632 & \nod & \nod  \\
8811.67 & \ion{Al}{2} 1670.787         & $1.880$               & 4.2740 & 2.55 &  13.7 \\
9131.04 & \ion{Mg}{2} 2796.352         & $0.612$               & 2.2653 & 1.74 &  13.6 \\
9155.12 & \ion{Mg}{2} 2803.531         & $0.305$               & 2.2656 & 2.10 &  14.0 \\
9162.51 & ?                            & \nod                  & \nod   & \nod & \nod  \\
9189.78 & \ion{Ni}{2} 1741.553         & $4.270\times 10^{-2}$ & 4.2768 & 0.91 &  14.9
\enddata
\tablecomments{Absorption features identified in the spectrum of \grb.  
We do not include metal lines blueward of the Ly$\alpha$ absorption
since at the low resolution of our spectrum these lines are blended
with Ly$\alpha$ forest features.  The columns are (left to right): (i)
Observed wavelength, (ii) line identification, (iii) oscillator
strength \citep{pgw+03}, (iv) redshift of the line, (v) rest-frame 
equivalent width, and (vi) logarithm of the column density assuming the 
optically-thin case (Equation~\ref{eqn:thin}); in most cases this is a 
lower limit since the lines are genearlly saturated.  In the case of 
blended features we derive the column density by assuming that the total
equivalent width is due to the highest redshift feature (typically the
host galaxy).  $^a$ Absorption features are located in the
atmospheric B band; $^b$ Absorption features are located in the
atmospheric A band.}
\end{deluxetable}

\clearpage
\begin{deluxetable}{lll}
\tablecolumns{3}
\tabcolsep0.2in\footnotesize
\tablewidth{0pc}
\tablecaption{Column Densities of Ions in $z_1$
\label{tab:columns}}
\tablehead {
\colhead {Ion}          &
\colhead {${\rm log}\,N$} &
\colhead {[X/H]}        \\
\colhead {}             &
\colhead {(cm$^{-3}$)}  &
\colhead {}             
}
\startdata
\ion{H}{1}      & 22.05 & \nodata \\
\ion{C}{1}      & 14.6  & $-3.8$  \\
\ion{C}{2}      & 17.0  & $-1.4$  \\
\ion{C}{4}      & 17.1  & $-1.3$  \\
\ion{O}{1}      & 16.5  & $-2.2$  \\
\ion{Al}{2}     & 16.3  & $-0.3$  \\
\ion{Si}{2}     & 15.7  & $-1.8$  \\
\ion{Si}{2}$^*$ & 15.1  & $-2.4$  \\
\ion{Si}{4}     & 15.6  & $-2.0$  \\
\ion{S}{2}      & 16.1  & $-1.2$  \\
\ion{Fe}{2}     & 15.5  & $-2.0$  \\
\ion{Ni}{2}     & 14.6  & $-1.7$
\enddata
\tablecomments{Ionic column densities and abundances relative to 
the Solar values as derived from the curve-of-growth analysis
(Figure~\ref{fig:cog}).  These values are lower limits in most cases
given the low spectral resolution and the location of most lines on
the flat portion of the COG.  Thus, the metallicity as inferred from
the sulfur abundance is $\gtrsim 0.06$ $Z_\odot$.}
\end{deluxetable}

\clearpage

\clearpage
\begin{figure}
\centerline{\psfig{file=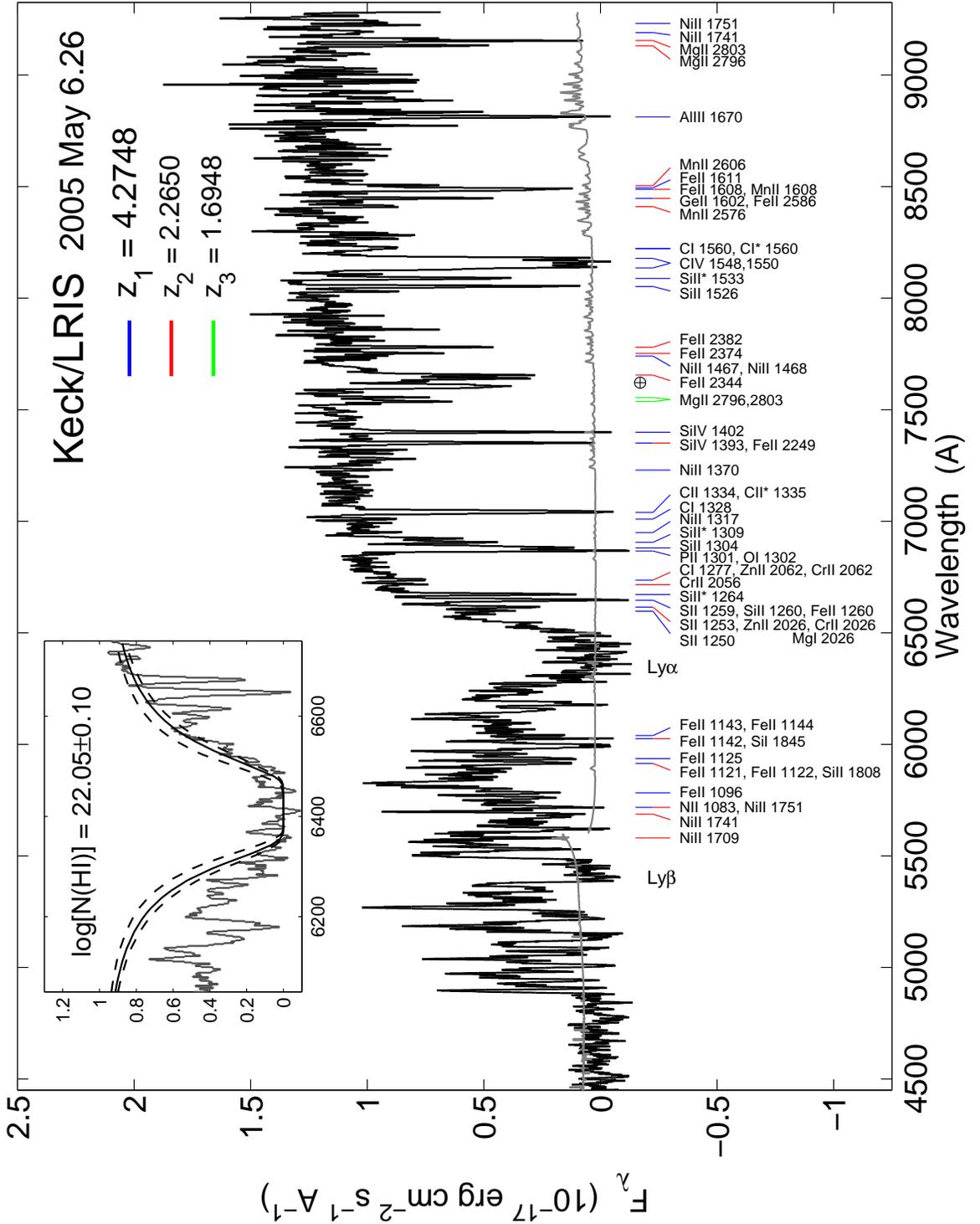,width=6.5in}}
\caption{Absorption spectrum of \grb\ obtained with LRIS on the Keck
I 10-m telescope.  Observational details are given in
\S\ref{sec:obs}.  Metal absorption features from all three systems
are shown, including lines that are blends from both $z_1$ and $z_2$.
We note that absorption features blueward of the damped Lyman alpha
feature are strongly blended with the Lyman alpha forest at the
resolution of our spectrum and their position is shown only for
completeness.  The inset shows a zoom-in of the Lyman alpha
absorption.  The solid line is the best fit with ${\rm
log}\,N(HI)=22.05$, and the dashed lines designate the $1\sigma$
uncertainty of 0.1 dex.
\label{fig:flam}}
\end{figure}

\clearpage

\clearpage

\clearpage

\clearpage

\clearpage

\clearpage

\end{document}